\documentclass[fleqn,usenatbib]{mnras}

\usepackage{newtxtext,newtxmath}

\usepackage[T1]{fontenc}

\DeclareRobustCommand{\VAN}[3]{#2}
\let\VANthebibliography\thebibliography
\def\thebibliography{\DeclareRobustCommand{\VAN}[3]{##3}\VANthebibliography}

\usepackage{graphicx}	
\usepackage{amsmath}	
\usepackage{array}

\usepackage{xcolor}
\usepackage{siunitx}
\usepackage{layouts}

\usepackage{booktabs}
\usepackage[export]{adjustbox}

\usepackage{lscape}

\definecolor{todo}{rgb}{0.9, 0.1, 0.1}
\definecolor{comment}{rgb}{0.2, 0.7, 0.3}
\definecolor{cite}{rgb}{0.2, 0.7, 0.7}

\DeclareSIUnit[]{\year}{yr}
\DeclareSIUnit[]{\parsec}{pc}
\DeclareSIUnit[]{\dex}{dex}
\DeclareSIUnit[]{\arcseconds}{as}
\DeclareSIUnit\solarmass{\ensuremath{\mathrm{M}_\odot}}
\DeclareSIUnit\solarradius{\ensuremath{\mathrm{R}_\odot}}
\DeclareSIUnit\solarlum{\ensuremath{\mathrm{L}_\odot}}

\title[Asteroseismic ages from \textit{Kepler}, K2 and TESS]{Asteroseismic ages for 17,000 stars in \textit{Kepler}, K2 and TESS}

\author[E. Willett et al.]{%
Emma Willett,$^{1}$\thanks{E-mail: e.m.willett.bham.ac.uk}
Andrea Miglio,$^{2, 3, 1}$
Saniya Khan,$^{4}$
Yvonne Elsworth,$^{1}$
Benoît Mosser,$^{5}$
\newauthor
Karsten Brogaard,$^{2, 6}$
Giada Casali,$^{7, 8, 2, 3}$
Cristina Chiappini,$^{9}$
Valeria Grisoni,$^{10, 2}$
Amalie Stokholm$,^{1, 2, 3, 6}$
\newauthor
Diego Bossini$^{11}$
and {William J. Chaplin}$^{1}$
\\
$^{1}$ School of Physics and Astronomy, University of Birmingham, Edgbaston, Birmingham, B15 2TT, UK\\
$^{2}$ Dipartimento di Fisica e Astronomia, Università degli Studi di Bologna, Via Gobetti 93/2, I-40129 Bologna, Italy\\
$^{3}$ INAF - Osservatorio di Astrofisica e Scienza dello Spazio di Bologna, Via Gobetti 93/3, I-40129 Bologna, Italy\\
$^{4}$ Institute of Physics, \'Ecole Polytechnique F\'ed\'erale de Lausanne (EPFL), Observatoire de Sauverny, 1290 Versoix, Switzerland\\
$^{5}$ LIRA, Observatoire de Paris, Universit\'e PSL, CNRS, Sorbonne Universit\'e, Universit\'e Paris Cit\'e, CY Cergy Paris Universit\'e, 92190 Meudon, France\\
$^{6}$ Stellar Astrophysics Centre, Department of Physics \& Astronomy, Aarhus University, Ny Munkegade 120, 8000 Aarhus C, Denmark\\
$^{7}$ Research School of Astronomy \& Astrophysics, Australian National University, Cotter Rd., Weston, ACT 2611, Australia\\
$^{8}$ ARC Centre of Excellence for All Sky Astrophysics in 3 Dimensions (ASTRO 3D), Stromlo, Australia\\
$^{9}$ Leibniz-Institut fur Astrophysik Potsdam (AIP), An der Sternwarte 16, D-14482 Potsdam, Germany\\
$^{10}$ INAF, Osservatorio Astronomico di Trieste, via G.B. Tiepolo 11, I-34131, Trieste, Italy\\
$^{11}$ Department of Physics and Astronomy G. Galilei, University of Padova, Vicolo dell'Osservatorio 3, I-35122, Padova, Italy
}

\date{Accepted XXX. Received YYY; in original form ZZZ}

\pubyear{2023}

\begin{document}
\label{firstpage}
\pagerange{\pageref{firstpage}--\pageref{lastpage}}
\maketitle

\begin{abstract}
The availability of asteroseismic constraints for tens of thousands of red giant (RG) stars has opened the door to robust age estimates, enabling time-resolved studies of different populations of stars in the Milky Way.
This study leverages data from \textit{Kepler}, K2, and TESS, in conjunction with astrometric data from \textit{Gaia} DR3 and spectroscopic constraints from APOGEE DR17 and GALAH DR3, to infer parameters for over 17,000 RGs. 
We use the code PARAM to homogeneously infer stellar properties considering in detail the sensitivity of our results to different choices of observational constraints.
We focus on age estimation, identifying potentially unreliable age determinations, and highlight stars with unreliable $\Delta\nu$ measurements based on comparisons using \textit{Gaia} luminosities. These are particularly relevant in K2 data due to the short duration of the observations of each campaign, and therefore important to characterise for Galactic archaeology studies where the spatial range of K2 is a benefit.
Thanks to the combination of data from different missions we explore trends in age, mass, and orbital parameters such as $R_\mathrm{g}$ and $Z_\mathrm{max}$, and examine time-resolved [$\alpha$/M]-[Fe/H] planes across different Galactic regions. Additionally, we compare age distributions in low- and high-$\alpha$ populations and chemically selected \textit{ex situ} stars. The study also extends known mass-[C/N] ratio relationships to lower masses. The catalogues resulting from this work will be instrumental in addressing key questions in Galactic archaeology and stellar evolution, and to improve training sets for machine-learning-based age estimations.
\end{abstract}

\begin{keywords}
asteroseismology -- Galaxy: abundances -- Galaxy: evolution -- Galaxy: stellar content -- stars: abundances -- stars: kinematics and dynamics
\end{keywords}



\section{Introduction and motivation}
\label{sec:intro}

Long-duration, space-based photometry has provided precise asteroseismic constraints for tens of thousands of red giant (RG) stars. These constraints, when combined with measurements of effective temperature and metallicity, have paved the way for robust estimates of stellar properties - including the mass, which is key for understanding stellar evolution, Galactic stellar structures and exoplanet systems. As a result of the tight correlation between initial stellar mass and main sequence (MS) lifetime, we can now obtain age estimates for large samples of RG field stars. These reliable stellar masses and ages throw light on open questions of stellar structure and evolution as well as providing strong tests of stellar models \citep[see e.g.][for reviews]{2013ARA&A..51..353C, 2021RvMP...93a5001A}. In addition, robust age estimates combined with spectroscopic and astrometric information are opening the door to understanding the formation and evolution of the Milky Way (MW) through Galactic archaeology \citep[e.g., see][]{2002ARA&A..40..487F, 2021A&ARv..29....5M}.

For meaningful studies of our Galaxy, we require samples which reach different Galactic domains, extending as far as possible both spatially and temporally. Low-mass RGs are excellent candidates for this work, being intrinsically luminous and long-lived, they can be seen over several kpc and probe more than \qty[]{10}{\giga\year}{} of MW history. Asteroseismic constraints have been used in this context in the Galactic disc \citep[e.g.][]{2013EPJWC..4303004M, 2016MNRAS.455..987C, 2017A&A...600A..70A, 2018MNRAS.475.5487S, 2019MNRAS.490.4465R, 2019MNRAS.490.5335S, 2021A&A...645A..85M, 2023MNRAS.524.1634S, 2023MNRAS.526.2141W, 2024AJ....167...50S, 2024A&A...685A.150V, 2024AJ....167..208W} and the halo \citep[e.g.][]{2019A&A...627A.173V, 2020NatAs...4..382C, 2021ApJ...916...88G, 2021ApJ...912...72M, 2021NatAs...5..640M, 2022MNRAS.514.2527B}. Ages from asteroseismology have also been used to verify other age-estimation techniques, such as chemical clocks and gyrochronology of main sequence stars \citep[e.g.][]{2021NatAs...5..707H, 2021A&A...646A..78M, 2022A&A...660A..15M}, and as training data for machine learning techniques \citep{2019MNRAS.489..176M, 2021MNRAS.503.2814C, 2023A&A...678A.158A, 2023MNRAS.522.4577L}. The latter, in particular, represents a powerful way to vastly increase the sample of stars with age information available but relies on high-quality training data, and inferences made by extrapolating beyond the parameter space covered by the training set are unlikely to be reliable. For this reason, increasing the number and variety of stars with robust, homogeneously determined ages from more `direct' methods, like asteroseismology, will support efforts to answer open questions about the MW and the stars in it on a large scale.

In this work, we present reliable ages for over $17,000$ stars, based on \textit{Kepler}, K2 and TESS photometry, APOGEE and GALAH spectroscopy and \textit{Gaia} astrometry. We show that asteroseismic results from the three photometry missions are complementary, with each representing a different compromise between the precision of the observations and scale of the coverage. We provide stellar and orbital parameters which are homogeneously determined across the samples, and consider in detail the sensitivity of our results to different choices of observational constraints. We also explore the applicability of each set of parameters to different problems, showing that the evolutionary phase of the star can affect the reliability of its age estimate, and provide quality flags to combat this effect.
We use our samples to explore several well-known chrono-chemo-kinematic relations in Galactic archaeology studies, showing trends with age in kinematic and chemical properties which we find to be robust against the uncertainties in the stellar parameters. Such trends may be used to constrain the dynamical and chemical evolution of the MW, and the K2 dataset, including stellar and orbital parameters, presented in this work has already been utilised in this context to study the radial metallicity gradient of the thin disc \citep{2023MNRAS.526.2141W}, cerium enrichment \citep{2023A&A...677A..60C}, and young $\alpha$-rich stars \citep{2024A&A...683A.111G}, while the \textit{Kepler} and K2 samples have been used to investigate mass loss during the red-giant-branch phase \citep{2024A&A...691A.288B}. We also identify the stars in our sample which were likely born outside the MW and present some results on stellar evolution during the RG phase, thus addressing open topics in both Galactic archaeology and stellar physics.

This paper is organised as follows: in Section \ref{sec:obs} we present the observational data and describe the cross-matched samples, and we describe the inference of stellar and orbital parameters in Section \ref{sec:inf}. Section \ref{sec:app} presents some applications of our results, in both Galactic archaeology and stellar evolution. A brief summary is provided in Section \ref{sec:conclusion}.


\section{Observational constraints}
\label{sec:obs}
The catalogues compiled in this work are based upon asteroseismic observations from the \textit{Kepler} \citep{2010Sci...327..977B, 2010PASP..122..131G}, K2 \citep{2014PASP..126..398H} and TESS \citep{2015JATIS...1a4003R} missions. These are combined with astrometric parameters from \textit{Gaia} DR3 \citep{2016A&A...595A...1G, 2023A&A...674A...1G} and spectroscopically derived quantities from APOGEE DR17 \citep{2017AJ....154...94M, 2022ApJS..259...35A} and GALAH DR3 \citep{2015MNRAS.449.2604D, 2019MNRAS.490.5335S, 2021MNRAS.506..150B}. In Sections \ref{ssec:obs_seismo} - \ref{ssec:obs_spectro} we introduce the observations, required parameters and quality flags and cross-match procedure, and describe the final samples in Section \ref{ssec:obs_samples}.

\subsection{Asteroseismic constraints}
\label{ssec:obs_seismo}

In this work, we make use of the global asteroseismic parameters $\nu_\mathrm{max}$, the frequency of maximum oscillation power, and $\Delta\nu$, the average large frequency separation. These parameters are related to the surface gravity and mean stellar density, respectively, so can be used to infer the stellar mass and radius \citep[see][]{2013ARA&A..51..353C}. In the following sections, we outline some relevant details of each asteroseismic mission and how $\nu_\mathrm{max}$ and $\Delta\nu$ are obtained.

\subsubsection{\textit{Kepler}}

The NASA \textit{Kepler} mission provided \qty[]{4}{years} of observations from 2009 for a single \qty[]{105}{\deg\squared}{} field of view. Our sample is based on the catalogue of \citet{2018ApJS..236...42Y}, which contains more than 16,000 RGs. These stars were selected from six previously published catalogues, but the lightcurves were reanalysed to produce a set of homogeneously determined $\nu_\mathrm{max}$ and $\Delta\nu$ values which we adopt in this work. Evolutionary states are also provided for some stars and, where available, determine which models are used in PARAM (Section \ref{ssec:inf_PARAM}). To this catalogue, we add the 2MASS ID and magnitudes available in the \textit{Kepler} Input Catalogue \citep[KIC,][]{2011AJ....142..112B}.

\subsubsection{K2}

Following the failure of \textit{Kepler}'s second reaction wheel in 2013, the mission was repurposed as K2, which went on to observe 20 campaigns around the ecliptic plane, each of 80 days duration \citep[see][for a description of the selection function]{2022MNRAS.517.1970S}. Here we use lightcurves produced by the EVEREST pipeline \citep{2018AJ....156...99L}, analysed with two independent methods. The pipeline of \citet[][hereafter BHM]{2020RNAAS...4..177E}, uses $\nu_\mathrm{max}$ determined by the deep learning method of \citet{2018MNRAS.476.3233H, 2018ApJ...859...64H} as a soft prior, and then applies several layers of weak tests to exclude false results without removing real, but unusual, stars. \citet{2009A&A...508..877M} and \citet{2010A&A...517A..22M} provide an alternative pipeline (hereafter COR) and, in general, the results of these two approaches agree well (see Appendix \ref{app:pipeline}). In the interest of clarity, the results presented here focus on the BHM pipeline, as \citet{2021MNRAS.502.1947M} found that it produces fewer false positive detections and the definition of $\Delta\nu$ is more similar to that used in our models. We add the 2MASS ID and magnitudes from the Ecliptic Plane Input Catalogue \citep[EPIC,][]{2016ApJS..224....2H}, removing cases where a single 2MASS ID is mapped to multiple K2 IDs. Because some targets were observed in multiple K2 campaigns, we also add a flag to identify the observation to be retained in our analysis, to prevent double counting when utilising our samples. We retain targets with a higher SNR on $\nu_\mathrm{max}$, meaning that the flag may be different for the BHM and COR pipelines.

\subsubsection{TESS}

TESS was launched in 2018 and will survey the whole sky with repeated \qty[]{27}{day} observations of $96 \times \qty[]{24}{\deg}$ sectors. In this work, we use on the results from the southern continuous viewing zone (SCVZ), where these sectors overlap, from the first year of observations. \citet{2021MNRAS.502.1947M} extracted lightcurves from the full frame images, which were analysed by three asteroseismic pipelines. Here, we focus on the results from the BHM pipeline.

\subsubsection{Cuts in asteroseismic parameters}
\label{sec:seismo_cuts}

We apply two cuts in the asteroseismic observables. First, we remove targets with $\nu_\mathrm{max}$ more than three standard deviations below \qty[]{20}{\micro\hertz}{}, to mitigate contamination by early-AGB stars where we cannot robustly determine the age, and as asteroseismic inference based on global asteroseismic parameters has not undergone the same robust testing as for less evolved targets. It should be noted that this cut also removes some evolved Red Giant Branch (RGB) targets, which would be an important consideration in the comparison of age distributions drawn from these observations and models of the chemo-dynamical evolution of the Milky Way. Since that is not the focus of this work, we do not address it further here. Second, we remove targets where $\nu_\mathrm{max}$ may be affected by the Nyquist frequency of the observations. Such cases are identified by comparing $\nu_\mathrm{max}$ and $\Delta\nu$ and removing targets where the tight correlation between these quantities is broken -- i.e. where $\Delta\nu$ is too large for the measured $\nu_\mathrm{max}$. In practice, this means removing targets with $\Delta\nu \geq \qty[]{21}{\micro\hertz}{}$ (see Appendix \ref{app:nyquist}). 

\subsection{Astrometric constraints}
\label{ssec:obs_gaia}

We obtain the required astrometric constraints from \textit{Gaia} DR3, which is based on \qty[]{34}{months} of observations. We use five-parameter solutions $[\alpha, \delta, \varpi, \mu_\alpha, \mu_\delta]$ and, following \citet{LL:LL-124}, remove targets with \texttt{ruwe} $> 1.4$ or which are marked as binaries by the \texttt{non\_single\_star} flag. We also consider different corrections to the known zero-point offset in \textit{Gaia} parallaxes \citep{2021A&A...649A...4L}. \citet{2023A&A...677A..21K} compared the \textit{Gaia} parallaxes with an independent estimate based on asteroseismology and found that the correction of \citet{2021A&A...649A...4L} works well for the \textit{Kepler} field but significantly overestimates the zero-point offset in K2 (averaged over multiple campaigns) and underestimates it in the TESS SCVZ. In addition, they also found a significant trend in parallax difference with stellar mass in some K2 campaigns. Though the underlying cause is not clear, it may be related to the different noise levels of the K2 campaigns, the variation in the quality of the Gaia data as a result of the scanning law, and the accuracy of the seismic parameters. For this reason, in this work, we apply the \citet{2021A&A...649A...4L} correction to the \textit{Kepler} and TESS sample, but use a simpler constant correction of \qty[]{17}{\micro\arcseconds}{} for K2 (guided by the average offset found in quasars in DR3 by \citealt{2021A&A...649A...4L}, and the results of \citealt{2023A&A...677A..21K}). We test the sensitivity of our results to the zero-point correction in Section \ref{ssec:inf_PARAM}. 

For the \textit{Kepler} and K2 samples, we obtain the \textit{Gaia} DR3 \texttt{source\_id} via crossmatch to the \texttt{TMASS\_PSC\_XSC\_BEST\_NEIGHBOUR} table \citep{2022gdr3.reptE..15M}, removing targets with multiple matches. 
The TESS SCVZ catalogue of \citet{2021MNRAS.502.1947M} already contains the DR2 \texttt{source\_id} so we use the \texttt{DR2\_NEIGHBOURHOOD} table instead \citep{2022gdr3.reptE..16M}. We retained sources with smaller angular separation in cases where a single DR2 \texttt{source\_id} matches multiple DR3 results. Finally, for all samples, we obtain data from the DR3 \texttt{gaia\_source} table, by matching on the \texttt{source\_id}.


\subsection{Spectroscopic constraints}
\label{ssec:obs_spectro}

The main spectroscopically derived parameters required in this work are the radial velocity (RV), effective temperature and chemical abundances -- specifically the iron abundance\footnote{Throughout this work we use iron abundance as a tracer for metallicity: $\mathrm{\left[Fe/H\right]} = \log_{10}\left(N_\mathrm{Fe}/N_\mathrm{H}\right)_* - \log_{10}\left(N_\mathrm{Fe}/N_\mathrm{H}\right)_\odot$, where $N_\mathrm{Fe}$ and $N_\mathrm{H}$ are the numbers of iron and hydrogen nuclei per unit volume of the stellar photosphere, respectively.} and average $\alpha$-element abundance.

\subsubsection{APOGEE}
\label{sssec:APG}

We cross-match each asteroseismic sample with APOGEE DR17, which uses high-resolution ($R \sim 22,500$) infra-red (IR) spectroscopy to obtain stellar parameters and abundances from 20 elements, covering a range of nucleosynthesis channels. Chemical abundances and stellar parameters are obtained from the APOGEE Stellar Parameters and Chemical Abundances Pipeline \citep[ASPCAP,][and see Holtzman et al. in preparation for a full description of this pipeline as applied to APOGEE DR17]{2016AJ....151..144G}. We remove targets with \texttt{STAR\_BAD} or \texttt{STAR\_WARN} set in \texttt{ASPCAPFLAG}, and those with any \texttt{RV\_FLAG} set. In addition, when using any of the individual abundances we remove targets with the relevant \texttt{ELEM\_FLAG} set. The uncertainties reported by ASPCAP reflect only the internal errors, so we enforce minimum values of $\sigma_\mathrm{[Fe/H]} = \qty[]{0.05}{\dex}$ and $\sigma_{T_\mathrm{eff}} = \qty[]{50}{\kelvin}$ during the inference of stellar parameters (see Section \ref{ssec:inf_PARAM}).

Before crossmatching to APOGEE DR17, we remove duplicated observations of targets observed in multiple fields, retaining those with higher [Fe/H] SNR. We then crossmatched based on the 2MASS ID (for TESS this was first obtained from \texttt{TMASS\_PSC\_XSC\_BEST\_NEIGHBOUR} table).

\subsubsection{GALAH}

We also cross-match the K2 samples to GALAH DR3 on the 2MASS ID, another high-resolution ($R \sim 28,000$) spectroscopic survey across the visible and IR. GALAH provides abundances for up to 30 elements per star \citep{2017MNRAS.464.1259K}, and we use the recommended quality flags, requiring that \texttt{snr\_c3\_iraf == 0}, \texttt{flap\_sp == 0} and \texttt{flag\_fe\_h == 0} for all targets, and \texttt{flag\_X\_fe == 0} for other abundances \texttt{X}. While GALAH DR4 is now available \citep{2025PASA...42...51B}, we report here the results for DR3 as they are investigated in other works where our APOGEE DR17 dataset is used \citep[e.g.][]{2023A&A...677A..21K, 2024A&A...683A.111G} and the aim of this paper is to provide the full catalogues on which these other works are based.

\subsection{Samples}
\label{ssec:obs_samples}

\begin{figure*}
    \centering
	\includegraphics[width=0.9\linewidth]{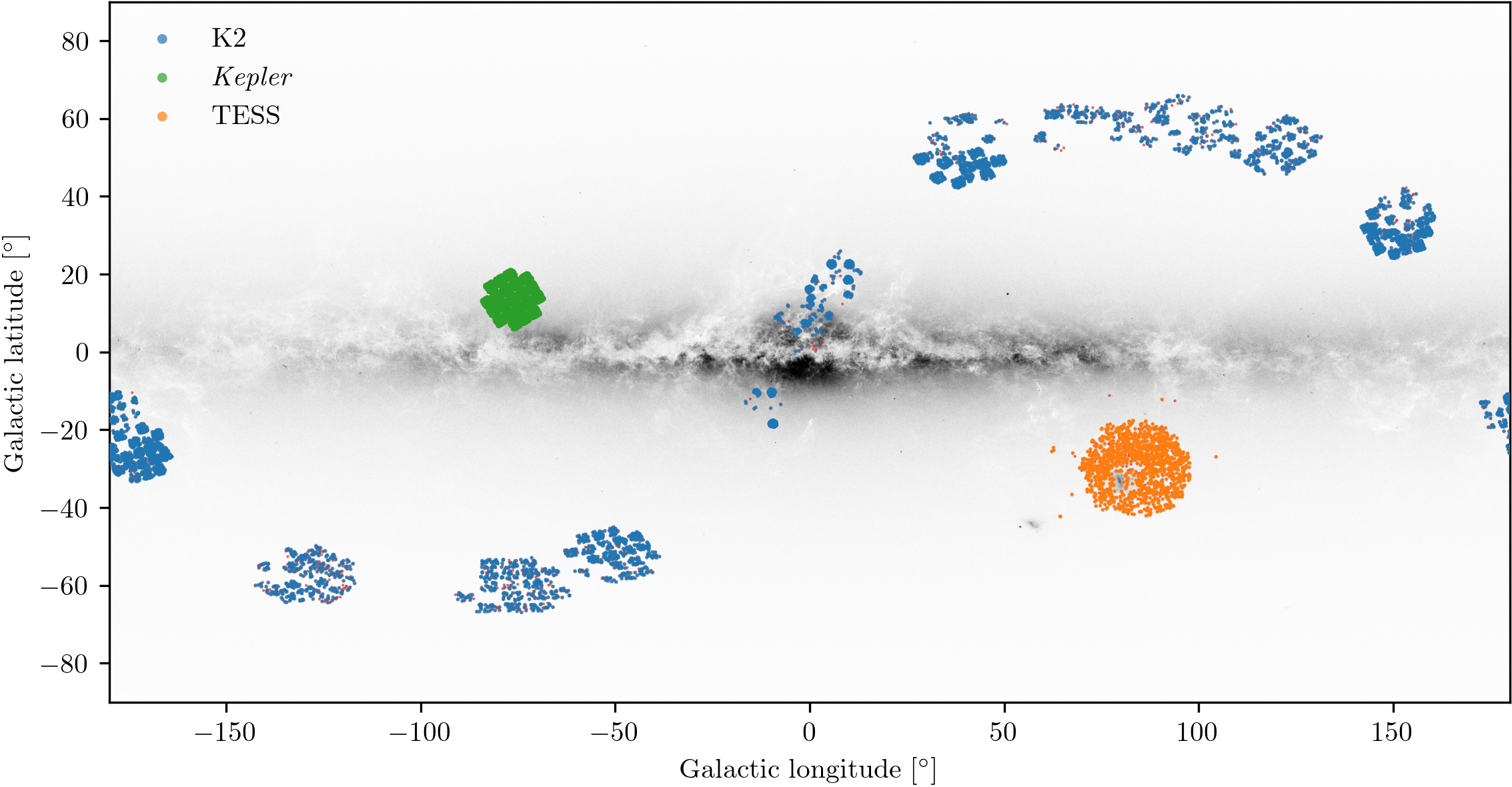}
    \caption{Location of stars in our samples in Galactic coordinates. Green, blue and orange points show stars with reliable ages from \textit{Kepler}, K2 and TESS, respectively. Stars with reliable data but ages which do not pass our tests are shown with red points. In this figure we use distances from PARAM and reliability tests based on $\Delta\nu$. This figure was generated using \texttt{mwplot}$^\mathrm{a}$, and the image in the background is modified from an image by ESA/Gaia/DPAC. \\
    \protect\footnotesize{$^\mathrm{a}$ \url{https://milkyway-plot.readthedocs.io/en/latest/index.html}}}
    \label{fig:sky_pos}
\end{figure*}

\begin{figure}
    \centering
	\includegraphics[width=\linewidth]{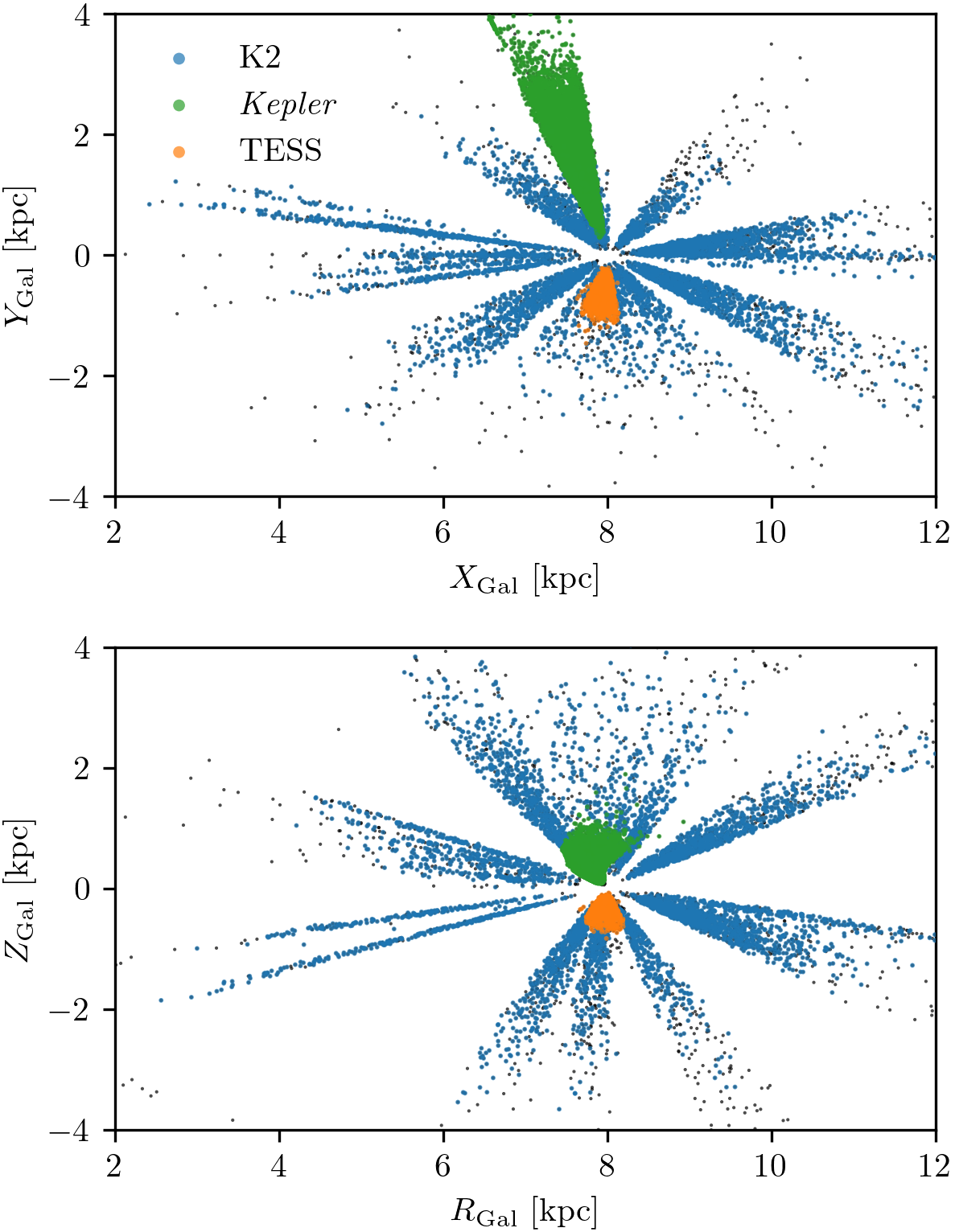}
    \caption{Location of stars in our sample in left-handed Galactocentric coordinates (top) and Galactocentric cylindrical coordinates (bottom). The colours are the same as Figure \ref{fig:sky_pos}. In this figure we use distances from PARAM and reliability tests based on $\Delta\nu$.}
    \label{fig:XY_RZ}
\end{figure}

\begin{figure*}
    \centering
	\includegraphics[width=0.9\linewidth]{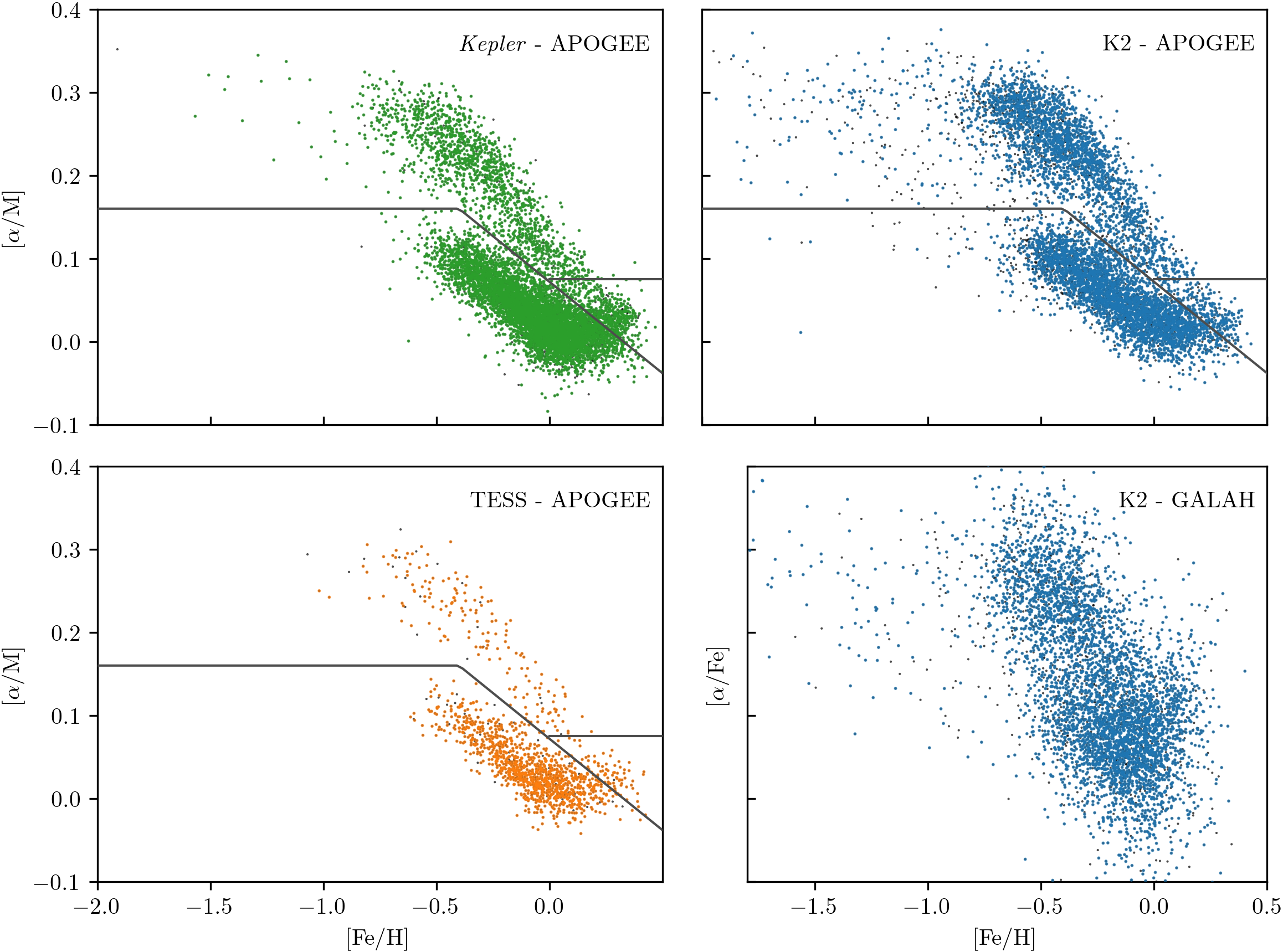}
    \caption{Average $\alpha$ element abundance [$\alpha$/M] for APOGEE samples or [$\alpha$/Fe] for the GALAH sample vs. [Fe/H]. The grey lines show the division between high-, low- and intermediate $\alpha$ populations used in Section \ref{sec:app_chemistry}. We have used reliability tests based on $\Delta\nu$ and the colours are the same as Figure \ref{fig:sky_pos}.}
    \label{fig:alpha_Fe}
\end{figure*}

In total, our samples contain observations of over $17,000$ stars. Figure \ref{fig:sky_pos} shows their positions in Galactic spherical coordinates, with the full sample shown in black and the stars with reliable ages (see Section \ref{ssec:inf_PARAM}) overlaid in colours corresponding to the survey. In Figure \ref{fig:XY_RZ} we show positions in Galactocentric Cartesian (top panel) and cylindrical coordinates (bottom panel). Finally, the samples are shown in APOGEE chemical abundance space using the well-studied [$\alpha$/M] vs. [Fe/H] plane\footnote{We use [$\alpha$/M] as a proxy for [$\alpha$/Fe] as the differences between them (defining [$\alpha$/Fe] as an average of [X/Fe] for O, Mg, Si, S, and Ca) are negligible.} in Figure \ref{fig:alpha_Fe}. For K2, we also show the GALAH [$\alpha$/Fe] vs. [Fe/H] plane, where the much greater scatter (expected due to the larger observational uncertainties) almost obscures the separation of the high- and low-$\alpha$ sequences.

\section{Inferring stellar and orbital parameters}
\label{sec:inf}

From the observations described above, we infer stellar (Section \ref{ssec:inf_PARAM}) and orbital (Section \ref{ssec:inf_orbits}) parameters for the stars in each sample. This section describes the assumptions made and, specifically in the case of the stellar masses and ages, the tests performed to ensure the most robust results are presented. In most cases, the figures shown in this section focus on the K2-APOGEE sample 
and additional figures corresponding to the \textit{Kepler} and TESS samples can be found in the Appendix, where relevant. The catalogues are available online, and described in Appendix \ref{app:Cats}.

\subsection{Stellar parameters}
\label{ssec:inf_PARAM}

Stellar masses ($M_*$), radii ($R_*$), ages ($\tau$), and distances ($D$) are inferred using the PARAM code \citep{2006A&A...458..609D, 2014MNRAS.445.2758R, 2017MNRAS.467.1433R}, which performs a Bayesian comparison between observations and a model grid. We follow the method presented in \citet{2021A&A...645A..85M} using the law of \citet{1975MSRSL...8..369R}, with efficiency parameter $\eta = 0.2$ to estimate the mass loss during the ascent of the RGB \citep{2012MNRAS.419.2077M}. We use the reference grid (G2) from \citet{2021A&A...645A..85M}, which was computed in MESA \citep{2011ApJS..192....3P, 2013ApJS..208....4P, 2015ApJS..220...15P, 2018ApJS..234...34P, 2019ApJS..243...10P} including microscopic diffusion and assuming a linear relation between the stellar helium and metal mass fraction ($\Delta Y/\Delta Z=1.3$) and the same relative surface effects as in the Solar model \citep[an alternative approach may be found in][]{2023MNRAS.523..916L}. Additional information about the model grid may be found in \citet{2017MNRAS.467.1433R}. For \textit{Kepler} stars with evolutionary state determinations available, either RGB or core-He burning models are considered by PARAM as appropriate. For \textit{Kepler} stars with unclassified evolutionary states, as well as stars from K2 and TESS, both sets of models are used. PARAM returns full posterior information, and we report the median and $16^\mathrm{th}$ and $84^\mathrm{th}$ percentiles here. We choose not to use evolutionary state information for the K2 and TESS samples because we consider this to be the more conservative approach. To require an evolutionary state would significantly reduce the number of stars in our samples, and the subset remaining may be subject to additional biases. Moreover, by keeping the evolutionary state as unknown, we account for the associated uncertainty in a statistically consistent way rather than potentially biasing the sample toward stars with confidently determined states.

Observational constraints are provided by $\nu_\mathrm{max}$, $T_\mathrm{eff}$, [Fe/H]\footnote{[Fe/H] is rescaled following the method of \citet{1993ApJ...414..580S} for $\alpha$-enhanced stars. See also the discussion in \citet{2021A&A...645A..85M}.} and we test using either $\Delta\nu$ or the luminosity, $L$, based on the (zero-point corrected) \textit{Gaia} parallax and 2MASS $K_s$ photometry \citep[][see Section \ref{sssec:Dnu_L}]{2006AJ....131.1163S}. An uncertainty floor of $\sigma_\mathrm{[Fe/H]} = \qty[]{0.05}{\dex}{}$ and $\sigma_{T_\mathrm{eff}} = \qty[]{50}{\kelvin}{}$ is imposed in the APOGEE observations, as the reported uncertainties reflect only the pipeline's internal errors. In the case of $T_\mathrm{eff}$, this has the additional benefit of reducing the impact of model uncertainties concerning the outer boundary conditions and near-surface convection. When using $L$ as a constraint, to further mitigate the impact of the uncertain models' ${T_\mathrm{eff}}$ scale on the inferred stellar parameters, we add the stellar mass estimated from the combination of the observed $\nu_\mathrm{max}$, $L$, and ${T_\mathrm{eff}}$ as an additional constraint (Section \ref{sssec:Dnu_L}). This approach has been tested using stellar models computed with different surface boundary conditions and ensures that the inferred $M_*$ is not significantly dependent on the models' ${T_\mathrm{eff}}$. The dependence of $M_*$ on ${T_\mathrm{eff}}$ is explored further in Section \ref{sssec:L_Teff}.

For \textit{Kepler} and K2 the luminosities were computed using extinctions from the \texttt{Bayestar19} dustmap \citep{2014ApJ...783..114G, 2019ApJ...887...93G} implemented in the \texttt{dustmaps} python package \citep{2018JOSS....3..695G}, from which we obtain quality flags for the convergence of the fit in each pixel and the reliability of the map at the queried distance. The TESS-SCVZ lies outside the range of \texttt{Bayestar19}, so we use the \texttt{combined19} dustmap \citep{2003A&A...409..205D, 2006A&A...453..635M, 2019ApJ...887...93G} implemented in \texttt{mwdust} \citep{2016ApJ...818..130B} instead. In both cases, we require that $\varpi / \sigma_\varpi \geq 5$ when using the luminosities or quantities derived from the luminosity, and few stars are also removed by the range of validity of the bolometric correction computation, performed using the code of \citet{2014MNRAS.444..392C, 2018MNRAS.479L.102C}.

As discussed in Section \ref{sec:intro}, several previous works have established ages for stars which overlap with those presented here. Of particular note, given the similar focus on applications to Galactic Archaeology, is the APO-K2 Catalogue of \citet{2024AJ....167..208W}. We present a brief comparison between the APO-K2 ages and those determined in this work in Appendix \ref{app:APO-K2_comp}.

\subsubsection{Using $\Delta\nu$ or $L$ as an observational constraint}
\label{sssec:Dnu_L}
\begin{figure}
    \centering
	\includegraphics[width=\linewidth]{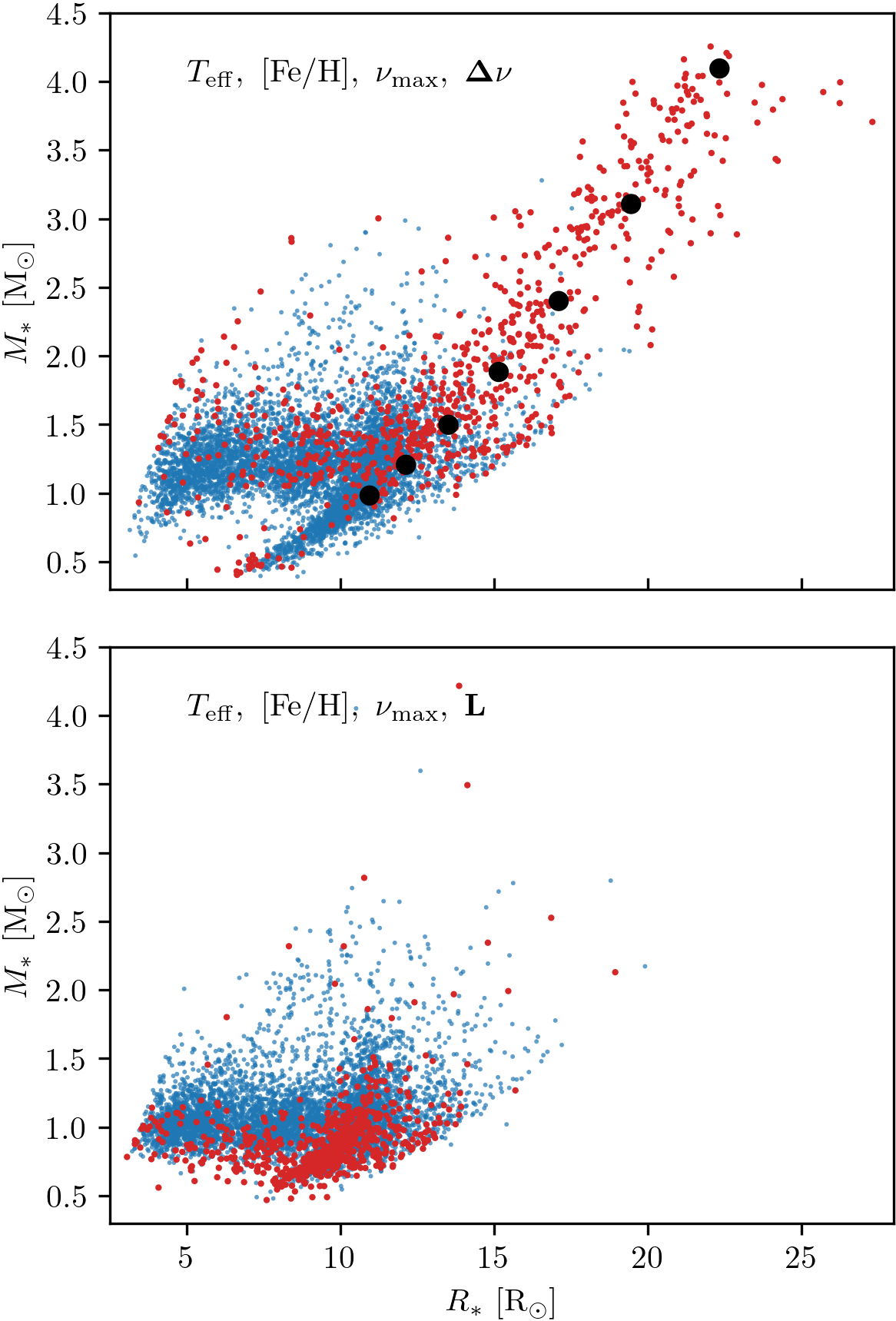}
    \caption{Stellar mass vs. radius for K2 stars from scaling relations where $\Delta\nu$ (top, Equations \ref{eq:M_Dnu} and \ref{eq:R_Dnu}) or $L$ (bottom, Equations \ref{eq:M_L} and \ref{eq:R_L}) are used as an observational constraint. Stars flagged for unreliable $\Delta\nu$ are shown in red (see Section \ref{sssec:Dnu_L} for a description of the flag). The large black points in the top panel show the effect of decreasing $\Delta\nu$ in steps of 5\% and recalculating  the mass and radius, for a star of initial mass $\qty[]{1}{\solarmass}{}{}$ and radius $\qty[]{11}{\solarradius}{}{}$. Stellar mass and radius increase with increasing bias on $\Delta\nu$, following the feature seen in the stars flagged for unreliable $\Delta\nu$.}
    \label{fig:M_R}
\end{figure}

In Figure \ref{fig:M_R} we show the stellar mass, $M_*$, and radius, $R_*$, obtained from scaling relations based on $T_\mathrm{eff}$, $\nu_\mathrm{max}$ and either $\Delta\nu$:
\begin{equation}
\label{eq:M_Dnu}
    \frac{M_*}{M_\odot} = \left(\frac{\nu_\mathrm{max}}{\nu_\mathrm{max,\,\odot}}\right)^3 \left(\frac{\Delta\nu}{\Delta\nu_\odot}\right)^{-4} \left(\frac{T_\mathrm{eff}}{T_\mathrm{eff,\,\odot}}\right)^\frac{3}{2}
\end{equation}
\begin{equation}
\label{eq:R_Dnu}
    \frac{R_*}{R_\odot} = \frac{\nu_\mathrm{max}}{\nu_\mathrm{max,\,\odot}} \left(\frac{\Delta\nu}{\Delta\nu_\odot}\right)^{-2} \left(\frac{T_\mathrm{eff}}{T_\mathrm{eff,\,\odot}}\right)^\frac{1}{2},
\end{equation}
or $L$:
\begin{equation}
\label{eq:M_L}
    \frac{M_*}{M_\odot} = \frac{\nu_\mathrm{max}}{\nu_\mathrm{max,\,\odot}} \frac{L}{L_\odot} \left(\frac{T_\mathrm{eff}}{T_\mathrm{eff,\,\odot}}\right)^\frac{-7}{2}
\end{equation}
\begin{equation}
\label{eq:R_L}
    \frac{R_*}{R_\odot} = \sqrt{\frac{L}{L_\odot} \left(\frac{T_\mathrm{eff}}{T_\mathrm{eff,\,\odot}}\right)^{-4}}.
\end{equation}

In the $\Delta\nu$ case (top panel) we see a strong feature with $M_* > \qty[]{1.5}{\solarmass}{}{}$ and $R_* > \qty[]{15}{\solarradius}{}{}$, which is not expected in descriptions of stellar evolution and is not present in the case where $\Delta\nu$ is replaced by $L$ (bottom panel). This is a consequence of the biases in $\nu_\mathrm{max}$ and $\Delta\nu$ which have previously been demonstrated in short-duration observations, such as K2 \citep[c.f. Equations \ref{eq:M_Dnu} and \ref{eq:R_Dnu} and see also][]{2022A&A...662L...7T}. We highlight stars which are common between the samples\footnote{There are some stars present in the $\Delta\nu$ sample which are removed when using $L$ as the observational constraint due to the quality flags on the extinction and parallax, described in Section \ref{ssec:inf_PARAM}.} where $|M_{*,\,\Delta\nu} - M_{*,\, L}| > 0.5 M_{*,\, L}$ (red points, masses computed from Equations \ref{eq:M_Dnu} and \ref{eq:M_L}). In the $\Delta\nu$ results, this highlights the high-$M_*$ - high-$R_*$ feature well, and from the $L$ results we see from the clustering around \qty[]{10}{\solarradius}{}{}, \qty[]{1}{\solarmass}{}{} that these stars are mostly members of the red clump (RC, and see also their position in the lower panel of Figure \ref{fig:cmd_flag}). This is an expected consequence of unreliable $\Delta\nu$ determinations for core helium burning (CHeB) stars with short-duration asteroseismic observations \citep{2022A&A...662L...7T}. Overlaid on the observations in large, black points is a simple demonstration of the effect of a bias in $\Delta\nu$. Starting from a star of $M_* = \qty[]{1}{\solarmass}{}{}$ and $R_* = \qty[]{11}{\solarradius}{}{}$, we decrease $\Delta\nu$ in steps of 5\% and recalculate the mass and radius from Equations \ref{eq:M_Dnu} and \ref{eq:R_Dnu}. The mass and radius increase as the bias on $\Delta\nu$ increases, following the high-$M_*$ - high-$R_*$ feature in the observations. In Appendix \ref{app:M_R_COR} we show that the situation is less significant when using the \textit{Kepler} and TESS SCVZ samples, where the observation durations are much longer than the K2 campaigns. We flag and remove stars with a high mass difference in samples where $\Delta\nu$ is an observational constraint, as their unreliably high masses will lead to erroneous, young ages.

\begin{figure}
    \centering
	\includegraphics[width=\linewidth]{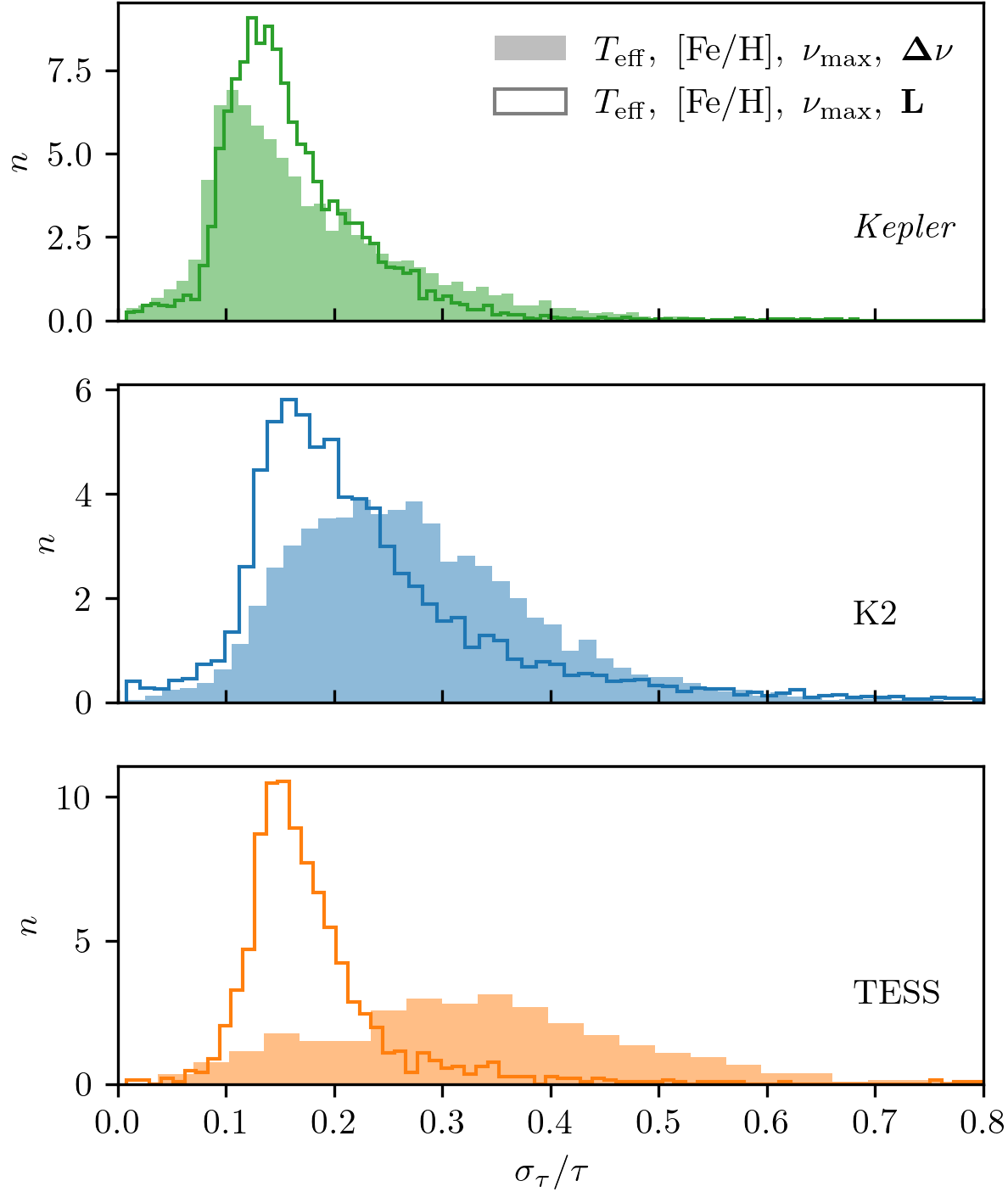}
    \caption{Distribution of the fractional uncertainty on age from PARAM where $\Delta\nu$ (filled) or $L$ (open) are used as an observational constraint for \textit{Kepler} (top), K2 (middle) and TESS (bottom). The area under each histogram integrates to one.}
    \label{fig:age_err}
\end{figure}

In the remaining sample, we find that ages obtained from PARAM and $\Delta\nu$ are less precise than those from $L$, thanks to the exquisite precision of the \textit{Gaia} DR3 data (Figure \ref{fig:age_err}). However, using $L$ as a constraint introduces other sensitivities in the results which are discussed further, below.

Table \ref{tab:num} shows a summary of the samples from different sets of observational constraints. The number of stars with reliable data is higher in the $\Delta\nu$ case because we do not require the additional checks on parallax uncertainty or dustmap flags. However, the number of stars with reliable ages in this case is smaller due to the removal of those with unreliable $\Delta\nu$ values.

\begin{table*}
\centering
\caption{Summary of reliable targets in datasets with different observational constraints. See text for descriptions of the criteria for reliable data and age, $\tau$. $L_{17}$ and $L_\mathrm{Ldg}$ indicate the luminosity derived from $\varpi + \qty[]{17}{\micro\arcseconds}{}$ or from $\varpi$ corrected according to the scheme of \citet{2021A&A...649A...4L}, respectively.}
\label{tab:num}
\begin{tabular}{llll|l|l}
\hline
Photometry source & Asteroseismology pipeline & Spectroscopy source & Observational constraint & $N_*$ with reliable data & $N_*$ with reliable $\tau$ \\ \hline
\textit{Kepler} & Yu 2018 & APOGEE DR17 & $\Delta\nu$                               & 8199 & 7873 \\
                &         &        & $L_\mathrm{Ldg}$                          & 8105 & 8005 \\ \hline
K2              & BHM     & APOGEE DR17 & $\Delta\nu$                               & 7390 & 6095 \\
                &         &        & $L_{17}$                                  & 6929 & 6831 \\
                &         &        & $L_\mathrm{Ldg}$                          & 7004 & 6976 \\ \cline{3-6} 
                &         & GALAH DR3  & $\Delta\nu$                               & 5266 & 4486 \\
                &         &        & $L_{17}$                                  & 5056 & 5041 \\ \cline{2-6} 
                & COR     & APOGEE DR17 & $\Delta\nu$                               & 7601 & 5751 \\
                &         &        & $L_{17}$                                  & 6883 & 6433 \\ \hline
TESS            & BHM     & APOGEE DR17 & $\Delta\nu$                               & 1398 & 1313 \\
                &         &        & $L_\mathrm{Ldg}$                          & 1398 & 1364 \\ \hline
\end{tabular}
\end{table*}

\subsubsection{Ages from $L$: sensitivity to $T_\mathrm{eff}$ and $\varpi$-zeropoint}
\label{sssec:L_Teff}

\begin{figure}
    \centering
	\includegraphics[width=\linewidth]{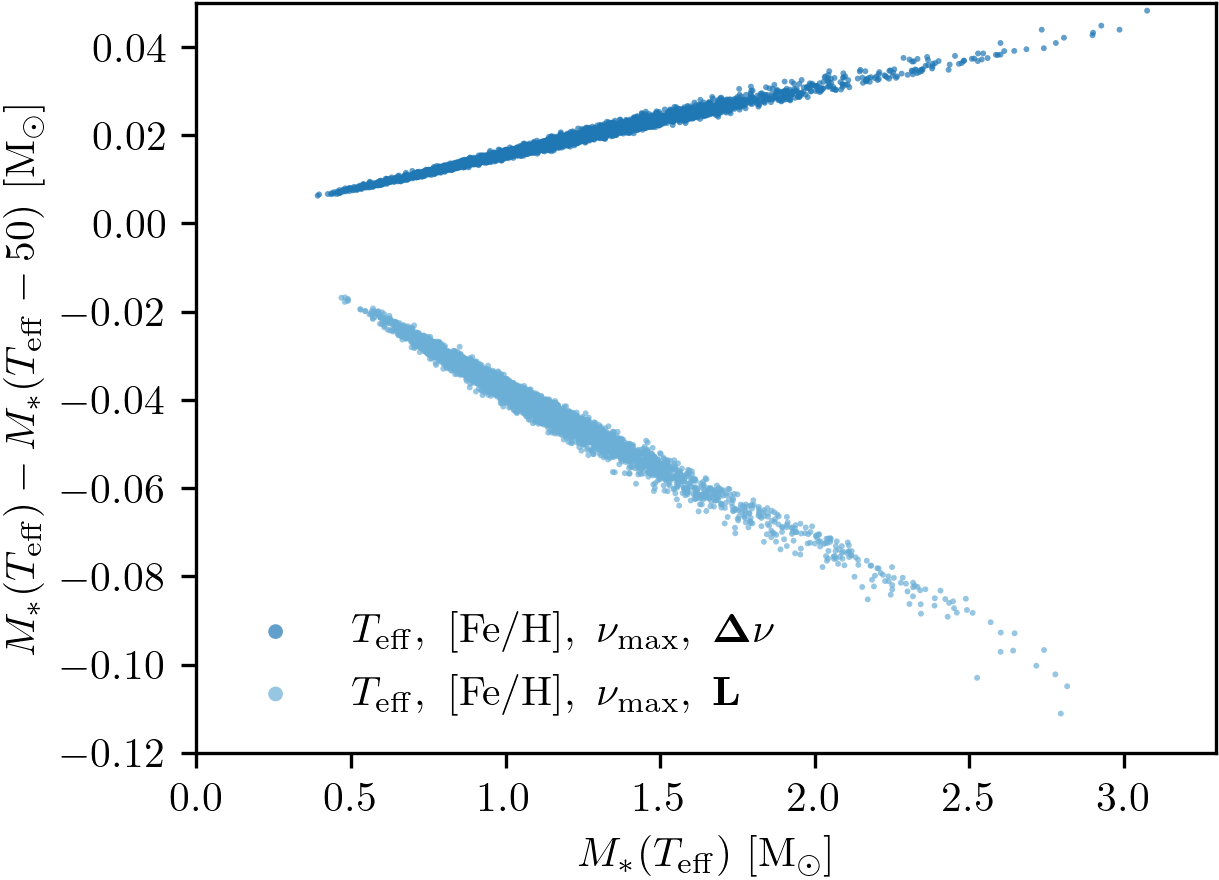}
    \caption{Difference between stellar masses for K2 stars obtained from scaling relations when $T_\mathrm{eff}$ is shifted by \qty[]{50}{\kelvin}{} vs. original mass. The cases where  $\Delta\nu$ (dark) or $L$ (light) are used as an observational constraint are shown.}
    \label{fig:sens_Teff}
\end{figure}

The mass, and therefore the age, from $L$ is much more sensitive to $T_\mathrm{eff}$, than when derived from $\Delta\nu$ (compare Equations \ref{eq:M_Dnu} and \ref{eq:M_L}). Given the known offsets in the temperature scales between spectroscopic surveys \citep[e.g.][]{2023A&A...670A.107H}, this sensitivity is a limitation of the ages constrained by $L$. The scale of this effect is shown in Figure \ref{fig:sens_Teff}, where we plot the mass difference from the scaling relations using $\Delta\nu$ (dark blue) or $L$ (light blue) caused by decreasing $T_\mathrm{eff}$ by \qty[]{50}{\kelvin}{}. For a \qty[]{1}{\solarmass}{}{} star, the mass difference when using $L$ is $\approx$ 5\%, corresponding to an age difference of more than \qty[]{1}{\giga\year}{}.

\begin{figure}
    \centering
	\includegraphics[width=\linewidth]{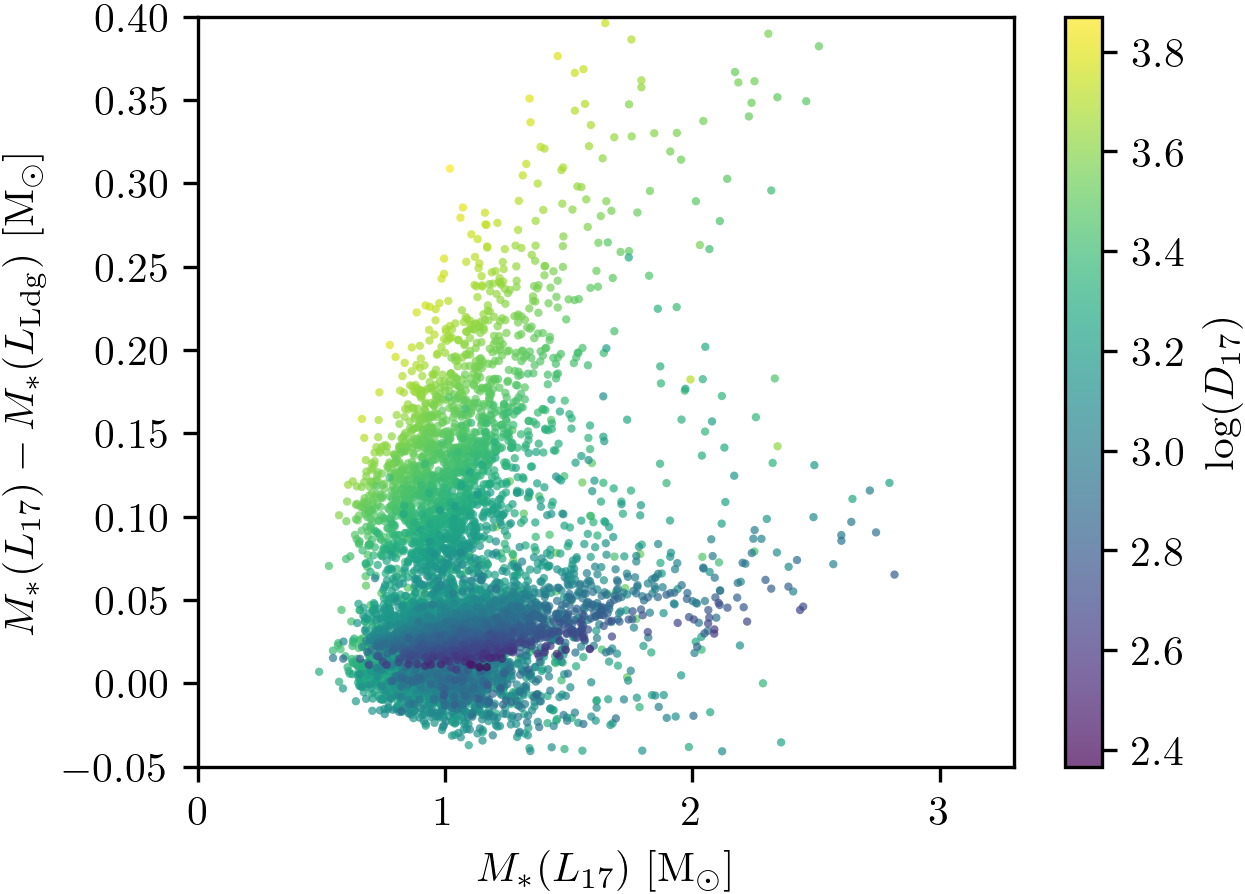}
    \caption{Difference between stellar masses for K2 stars obtained from scaling relations using luminosity when the correction to the \textit{Gaia} parallax zero-point offset is changed from a constant \qty[]{17}{\micro\arcseconds}{} ($L_{17}$) to the correction proposed by \citet{2021A&A...649A...4L} ($L_\mathrm{Ldg}$). The difference is shown as a function of the $L_{17}$ mass and coloured according to the distance from the inverse of the \textit{Gaia} parallax, with a constant zero-point correction of \qty[]{17}{\micro\arcseconds}{}.}
    \label{fig:sens_zpt}
\end{figure}

Using the luminosity also introduces a sensitivity to the zero-point correction to the \textit{Gaia} parallax. This is a further limitation and results in, effectively, data release-dependent ages. The situation is more complicated for K2, as \citet{2023A&A...677A..21K} found that the correction differs between K2 campaigns. In Figure \ref{fig:sens_zpt} we show the difference in mass, again from the scaling relation (Equation \ref{eq:M_L}) using the luminosity derived from $\varpi + \qty[]{17}{\micro\arcseconds}{}$, $L_{17}$, or with $\varpi$ corrected according to the scheme of \citet{2021A&A...649A...4L}, $L_\mathrm{Ldg}$. Stars are coloured according to the logarithmic distance and sorted with the closest stars on top. As expected, more distant stars are more sensitive to the zero-point correction, with a mass difference of more than 20\% in some cases.

\subsubsection{PARAM output: sensitivity to priors}
\label{sssec:priors}

\begin{figure}
    \centering
	\includegraphics[width=\linewidth]{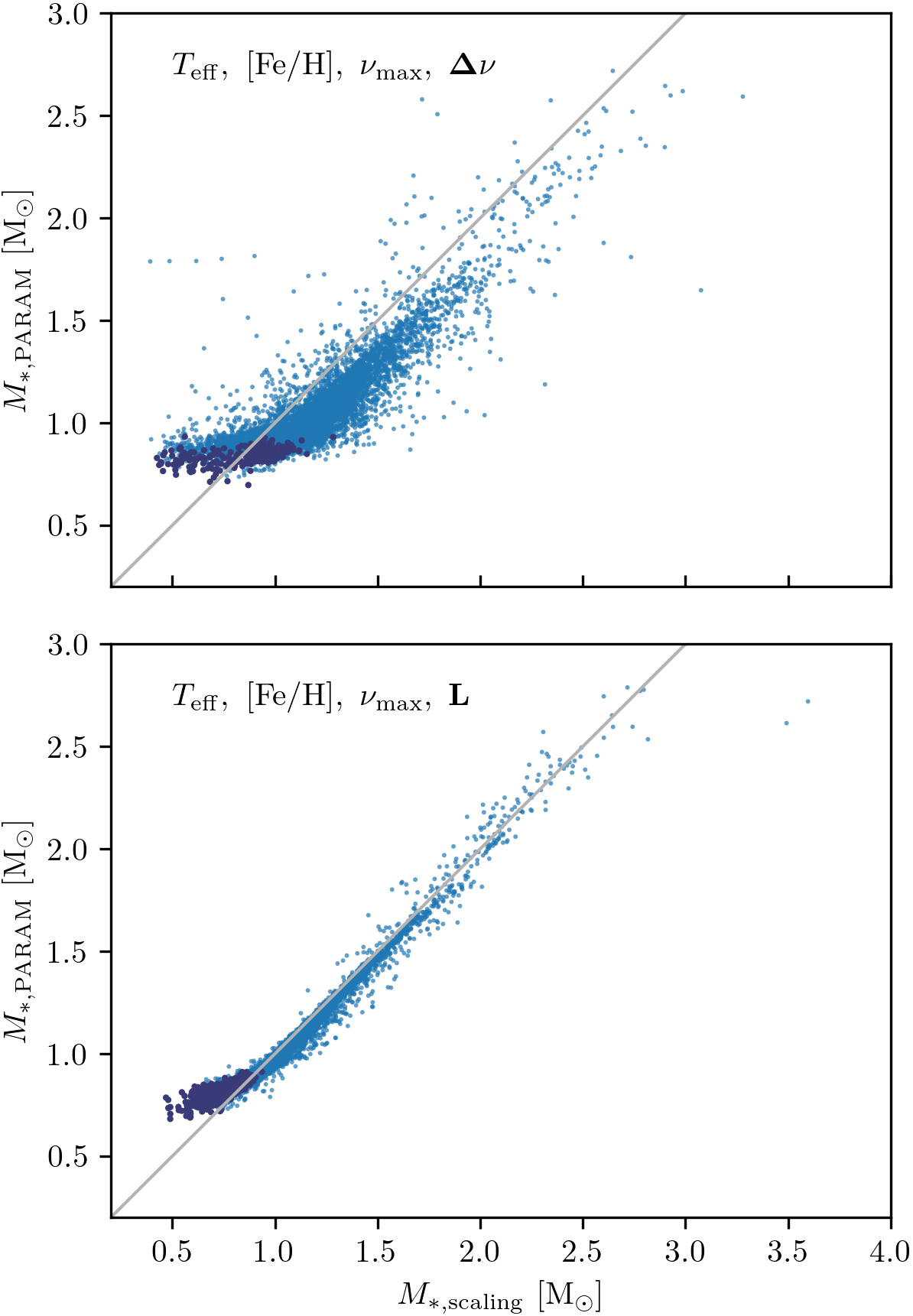}
    \caption{Stellar masses for K2 stars obtained from PARAM vs. those from scaling relations where $\Delta\nu$ (top) or $L$ (bottom) are used as an observational constraint. Stars with masses that may be dominated by the priors imposed in PARAM are shown in purple (see Section \ref{sssec:priors} for a description of the flag).}
    \label{fig:MParam_Mscaling}
\end{figure}

Regardless of the choice of observational constraint, when using the stellar parameters output by PARAM, it is important to understand any effects which may be introduced by the model grid or choice of priors. When using $[T_\mathrm{eff}, \mathrm{[Fe/H]}, \nu_\mathrm{max}, L]$ as observational constraints, the stellar mass from PARAM is inferred based on the scaling relation of the same quantities (Equation \ref{eq:M_L}), and any differences are therefore a result of the priors on the mass and age used in our analysis. The bottom panel of Figure \ref{fig:MParam_Mscaling} shows a comparison between the two where, at $M_\mathrm{scaling}$ below around $0.7 \mathrm{M}_\odot$, we see a deviation caused by the upper limit of the age prior used. We flag stars within $1\sigma$ of this limit (\qty[]{20}{\giga\year}{}), where $\sigma$ is the median fractional uncertainty of the whole population multiplied by the stellar age (purple points). We choose this estimate as the uncertainties reported on individual stars with posteriors affected by the prior will be artificially small, and this appears to select the affected stars well. We apply the same process to the results constrained by $[T_\mathrm{eff}, \mathrm{[Fe/H]}, \nu_\mathrm{max}, \Delta\nu]$, though the affected stars are not easily identifiable in a comparison of the masses (top panel) as PARAM uses $\langle\Delta\nu\rangle$ computed using the radial mode frequencies of the models in the grid, so the results are not directly comparable to the scaling relations in this case \citep[see][for details]{2017MNRAS.467.1433R}. The results constrained by $\Delta\nu$ are, therefore, an important improvement but, in both cases, the flagged stars are potentially unreliable, and we ensure that the results presented in the rest of this paper based on PARAM outputs are not affected by removing them.

\begin{figure}
    \centering
	\includegraphics[width=\linewidth]{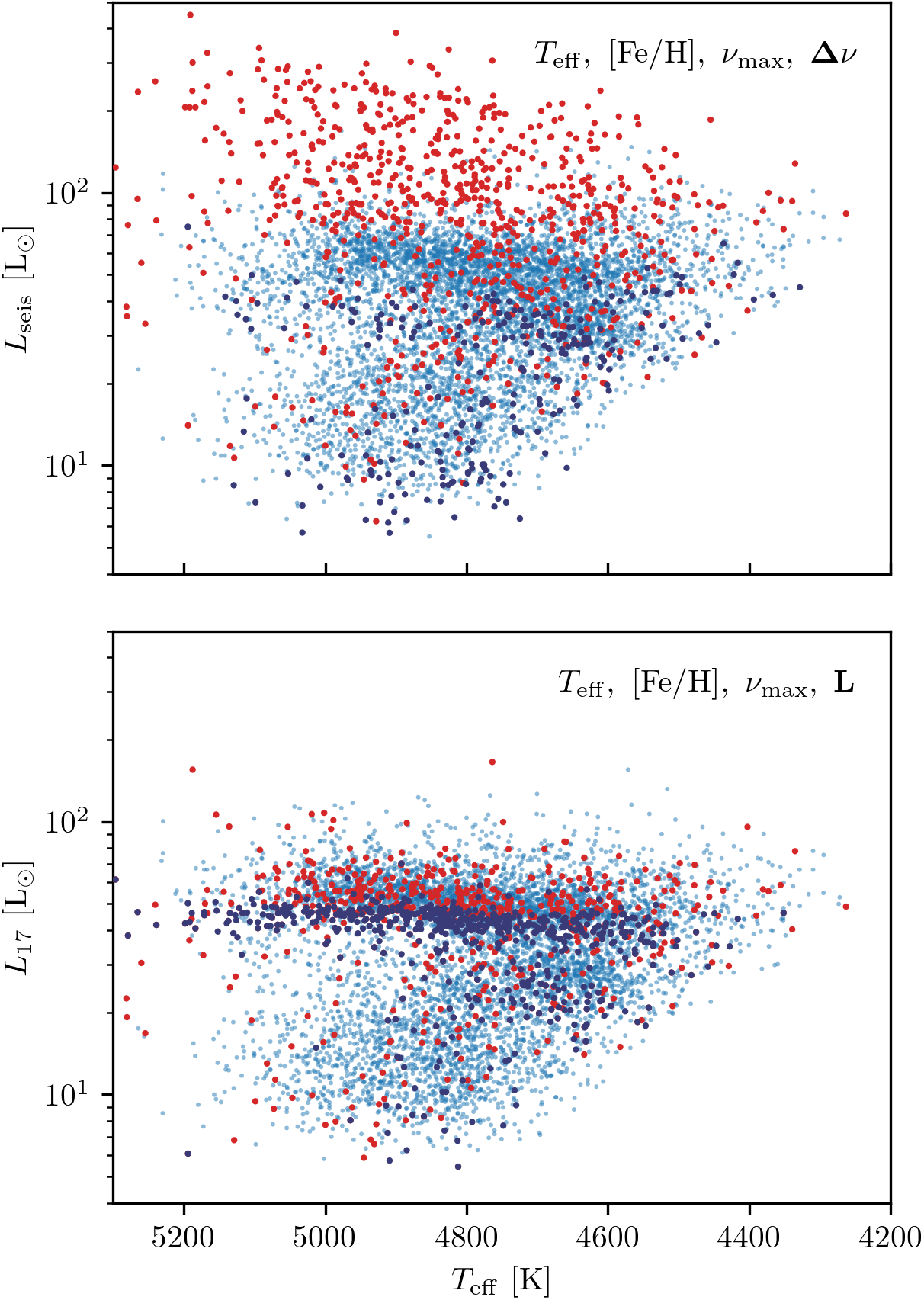}
    \caption{HRD showing K2 stars flagged for unreliable $\Delta\nu$ (red) and with masses that may be dominated by the priors imposed in PARAM are shown (purple; see Section \ref{ssec:inf_PARAM} for a description of the flags). The full sample is shown in the background (blue).}
    \label{fig:cmd_flag}
\end{figure}

With the possible exception of products of partial envelope stripping on the RGB \citep{2022NatAs...6..673L, 2023A&A...671A..53M, 2024A&A...691A..17M, 2025A&A...699A..39M}, the majority of stars at these low masses are CHeB (see also their position in the lower panel of Figure \ref{fig:cmd_flag}). These stars are likely to lose significant mass in the upper RGB \citep{2021A&A...645A..85M, 2021MNRAS.503..694T, 2024A&A...691A.288B, 2025ApJ...988..179L} and reach the CHeB phase with masses that would indicate, if mass-loss were not accounted for, ages much greater than the age of the Universe. Since our current prescription for mass loss is limited (see Section \ref{ssec:inf_PARAM}), we do not consider these stars to have robust ages and remove them from our sample. 

The top panel of Figure \ref{fig:cmd_flag} shows the Hertzsprung-Russell Diagram (HRD) for the sample with stellar parameters based on $\Delta\nu$ using the asteroseismic luminosity $L_\mathrm{seis}$,
\begin{equation}
\label{eq:L_seis}
    \frac{L_\mathrm{seis}}{L_\odot} = \left(\frac{\nu_\mathrm{max}}{\nu_\mathrm{max,\,\odot}}\right)^2\left(\frac{\Delta\nu}{\Delta\nu_\odot}\right)^{-4}\left(\frac{T_\mathrm{eff}}{T_\mathrm{eff,\,\odot}}\right)^5,
\end{equation}
and the bottom panel shows stars with parameters from $L_{17}$. The stars flagged as having unreliable $\Delta\nu$ values are shown in red and those with parameters possibly affected by the age prior are shown in purple. There are fewer prior-affected stars shown in the $\Delta\nu$ results as there is a significant overlap with those stars already removed by the $\Delta\nu$ flag. This means that the results based on $\Delta\nu$ are less sensitive to the prior, but have fewer CHeB stars overall.

\subsection{Orbital parameters}
\label{ssec:inf_orbits}

Orbital parameters are computed using the fast orbit estimation method of \citet{2018PASP..130k4501M}, implemented in \texttt{galpy} \citep{2015ApJS..216...29B} and a left-handed Galactocentric coordinate frame. We assume that the MW is well represented by the simple Milky Way potential \texttt{MWPotential2014} \citep{2015ApJS..216...29B}, and that the radial position of the Sun is $R_{\mathrm{Gal},\,\odot} = \qty[]{8.0}{\kilo\parsec}$, where the circular velocity is $v_\mathrm{circ} = \qty[]{220}{\kilo\meter\per\second}$ \citep{2012ApJ...759..131B}. The additional solar motion is given by $[U, V, W]_\odot = [-11.1, 12.24, 7.25]\qty[]{}{\kilo\meter\per\second}$ \citep{2010MNRAS.403.1829S}, and the Sun's vertical offset from the Galactic plane is $Z_{\mathrm{Gal},\,\odot} = \qty[]{20.8}{\parsec}$ \citep{2019MNRAS.482.1417B}.

To constrain the parameters and their uncertainties, we draw samples from the covariance matrix formed by $[\alpha, \delta, \mu_\alpha, \mu_\delta]$ (from \textit{Gaia}), RV (from APOGEE or GALAH) and either $D$ from PARAM, when using $\Delta\nu$ to constrain the stellar parameters, or \textit{Gaia} $\varpi$, when using $L$. In the latter case, we invert the sampled $\varpi$ to obtain an estimate for $D$. We assume that the spectroscopic RV and $D$ from PARAM are uncorrelated with each other and the other parameters, but in all other cases we utilise the correlation coefficients included in the \textit{Gaia} catalogue. The estimated parameters include the orbital energy, $E$, angular momentum, $L_Z$, guiding radius, $R_\mathrm{g}$, maximum vertical excursion, $Z_\mathrm{max}$, and eccentricity, $e$, and report the median and $16^\mathrm{th}$ and $84^\mathrm{th}$ percentiles. 

\section{Applications}
\label{sec:app}

In this section, we demonstrate some applications of the samples presented above. In all cases, we show only stars with reliable ages and, unless otherwise stated, use the stellar parameters from PARAM, constrained by $[T_\mathrm{eff}, \mathrm{[Fe/H]}, \nu_\mathrm{max}, \Delta\nu]$ and K2 results from the BHM-APOGEE sample. We choose to focus on the ages constrained by $\Delta\nu$ as, though they are less precise for K2 and TESS, the sensitivity to $T_\mathrm{eff}$ and $\varpi$-zeropoint is reduced (Section \ref{sssec:L_Teff}).

\subsection{Kinematic properties}
\label{sec:app_kinematic}

\begin{figure}
    \centering
	\includegraphics[width=\linewidth]{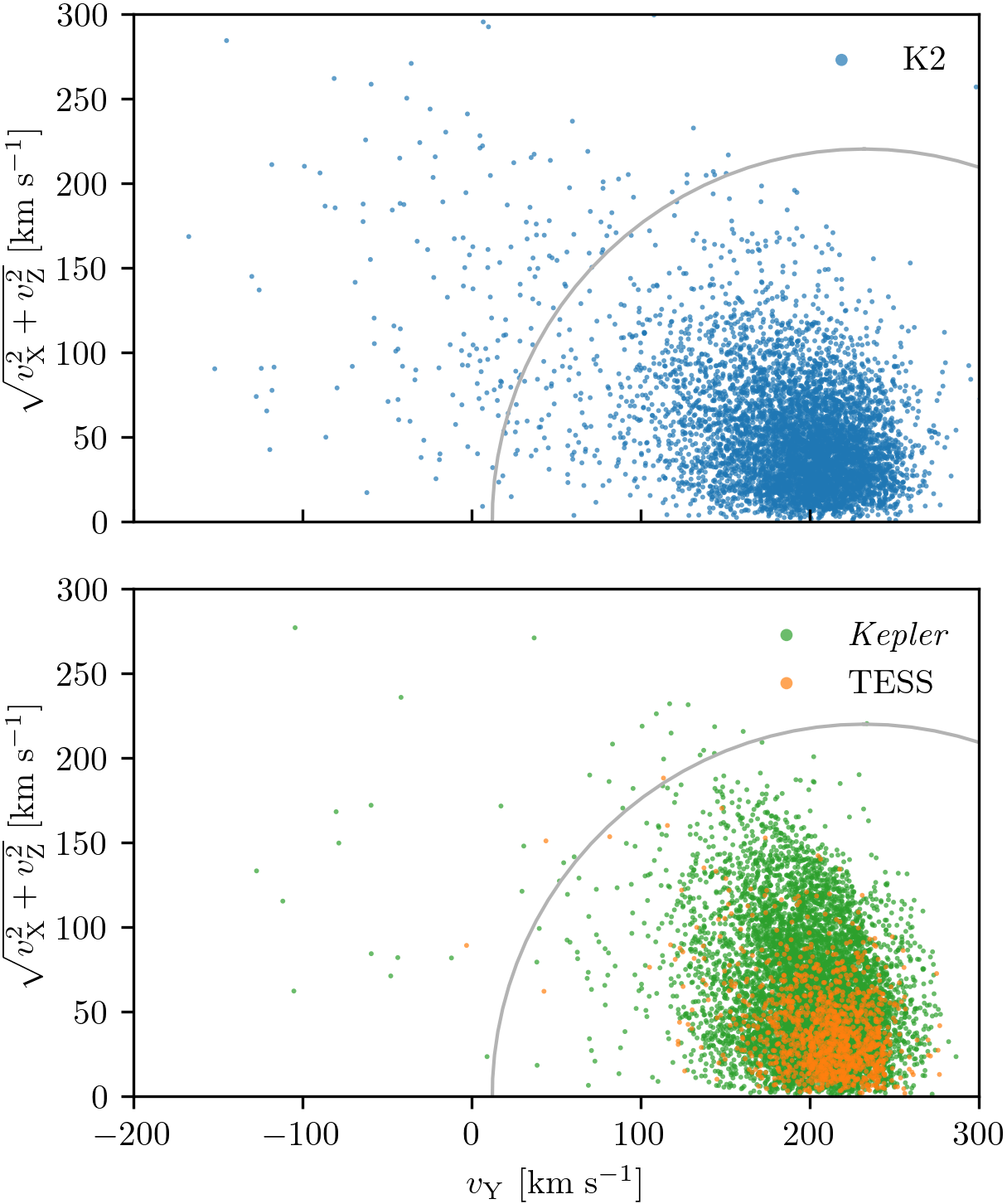}
    \caption{Toomre diagram for stars with reliable ages in the K2 (top), Kepler and TESS (bottom) samples. The grey line indicates the approximate separation of disc and halo stars and the colours are the same as Figure \ref{fig:sky_pos}.}
    \label{fig:Toomre}
\end{figure}

\begin{figure}
    \centering
	\includegraphics[width=\linewidth]{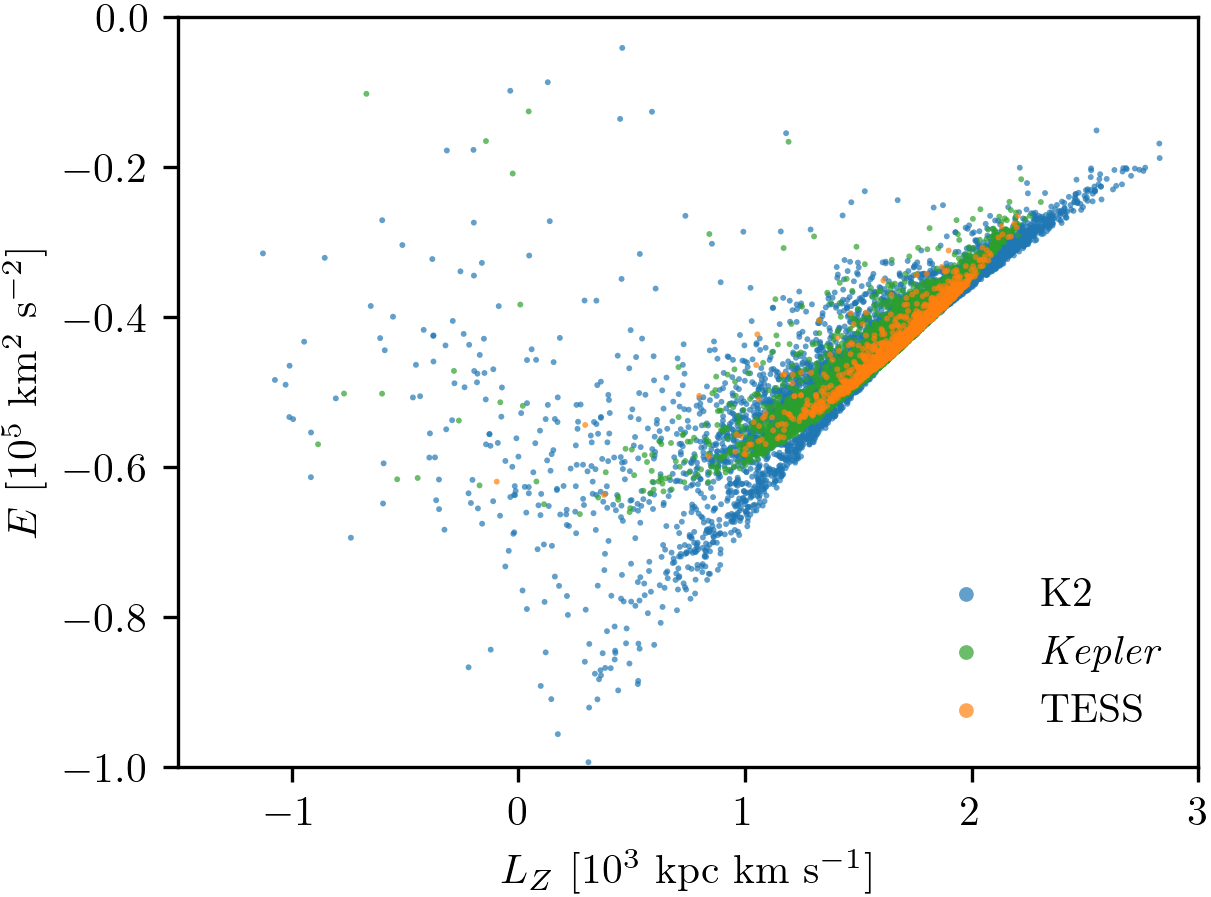}
    \caption{Energy vs. Z-component of angular momentum for stars with reliable ages. The colours are the same as Figure \ref{fig:sky_pos}.}
    \label{fig:ELz}
\end{figure}

First, in Figures \ref{fig:Toomre} and \ref{fig:ELz} we show two commonly used kinematic planes: the Toomre diagram (velocities in Galactocentric Cartesian coordinates) and orbital energy vs. angular momentum. In both planes, we see that the TESS sample is confined to the thin disc,  \textit{Kepler} extends into the thick disc and includes a few halo stars, and K2 provides better coverage of the halo (see Figure \ref{fig:Toomre}, where the grey line indicates the approximate separation of disc and halo stars). This is consistent with the chemical picture from Figure \ref{fig:alpha_Fe}, as the low- and high-$\alpha$ and low-metallicity ([Fe/H] $\lessapprox -0.7$) populations map well, in broad terms, to the thin and thick discs and halo, respectively. Figure \ref{fig:ELz} shows that the majority of stars on retrograde orbits which are present in our samples are found in K2, which also provides better coverage of the inner Galaxy (low-energy orbits).

\begin{figure}
    \centering
	\includegraphics[width=\linewidth]{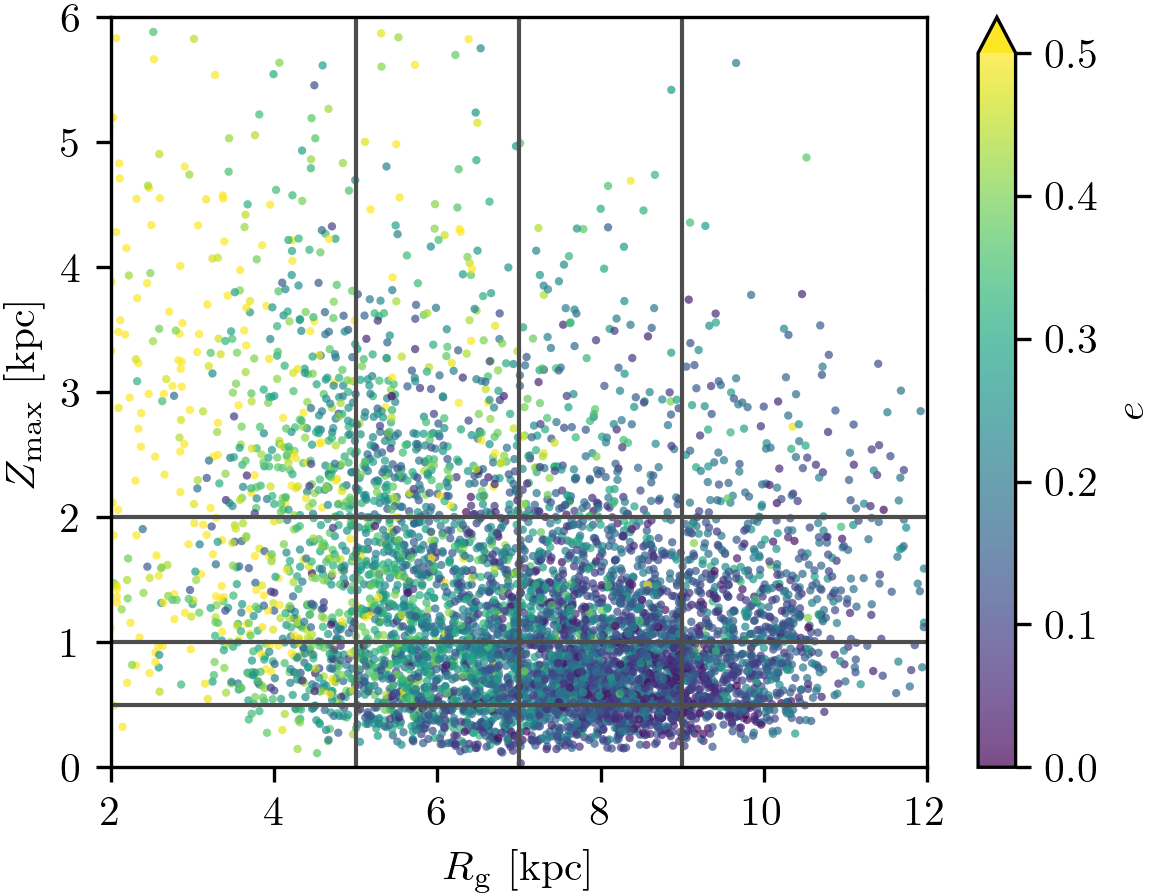}
    \caption{Maximum vertical excursion vs. guiding radius for K2 stars with reliable ages, coloured according to the eccentricity.}
    \label{fig:RgZmax}
\end{figure}

In many applications, it is useful to investigate spatial trends in the Galaxy and in these cases K2 is especially valuable (see Figure \ref{fig:XY_RZ} for a comparison of the samples' spatial distributions). However, using the present-day positions of stars does not provide the most representative view of the processes at work in the MW, particularly for older stars, as radial migration and vertical heating move stars away from their birth positions \citep[e.g.][]{2009MNRAS.396..203S, 2013A&A...558A...9M, 2014A&A...572A..92M, 2015ApJ...808..132H, 2018ApJ...865...96F, 2020ApJ...896...15F, 2020MNRAS.496...80V, 2021A&A...645A..85M}. This means that stars passing close to the Sun and within our observable limits today were not necessarily born nearby, and nor do they necessarily have orbits similar to the Sun. Figure \ref{fig:RgZmax} shows an alternative choice of coordinates, defined below.

The term radial migration is used to refer to two processes: blurring, the observational effect of orbital eccentricity, which results in the star being observed at different Galactocentric radii, $R_\mathrm{Gal}$, and churning, where interactions with non-axisymmetric features in the Galactic disc result in a change of the orbital radius. Both processes are more important in older populations, as orbital eccentricity tends to increase with time due to radial heating, and the interactions which lead to churning can occur at any time so older stars are more likely to have migrated. The guiding radius $R_\mathrm{g}$, is the radius of a circular orbit with the same $L_Z$ as the true orbit and therefore it mitigates the effect of blurring, providing a better estimate of the `average' radial position of the star. This has the effect of appearing to spread our sample further in radius, as many stars observed around $R_\mathrm{Gal} \approx \qty[]{8}{\kilo\parsec}{}$ spend the majority of their orbit closer to or further from the Galactic centre. This is particularly noticeable when comparing the number of stars with $R_\mathrm{Gal}$ and $R_\mathrm{g} \approx \qty[]{6}{\kilo\parsec}{}$ in Figures \ref{fig:XY_RZ} and \ref{fig:RgZmax}, where we see many more stars in the latter case, indicating that they are on more eccentric orbits and nearer to apocentre than pericentre.

For the vertical coordinate of Figure \ref{fig:RgZmax}, it is useful to know how far the orbit extends from the Galactic plane, as this is the main differentiating factor between orbits which are more thin or thick disc-like. The maximum vertical extent of the stellar orbit, $Z_\mathrm{max}$, is also expected to increase with time as stars' orbits are heated by adiabatic (due to the secular evolution of the Galactic disc) or non-adiabatic (due to interactions with other galaxies or giant molecular clouds) processes \citep{2019ApJ...878...21T}. Figure \ref{fig:RgZmax} shows the K2 sample divided into bins in $R_\mathrm{g}$ and $Z_\mathrm{max}$, which we use to investigate chemical trends later in this section.

\begin{figure*}
    \centering
	\includegraphics[width=0.7\linewidth]{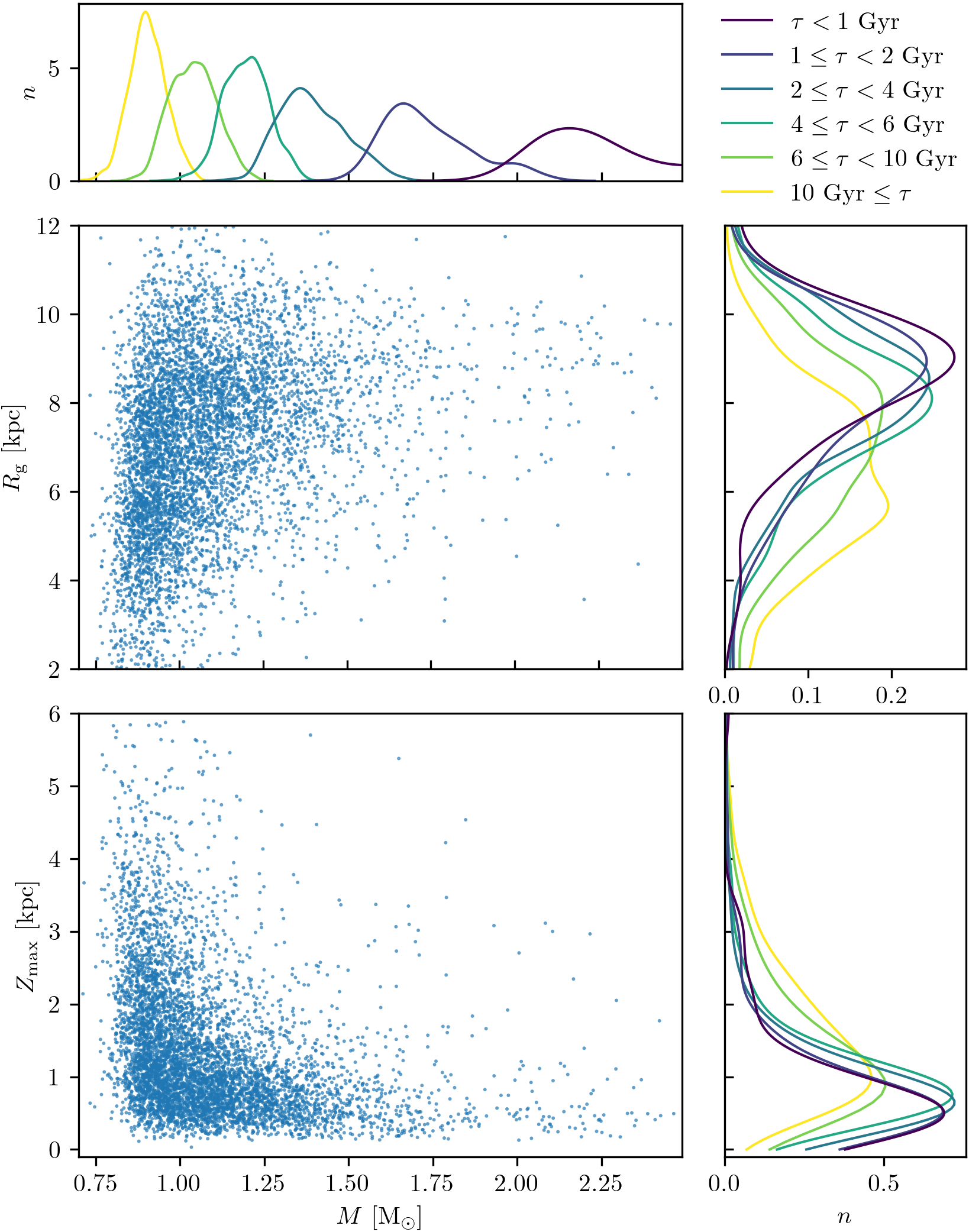}
    \caption{Guiding radius (middle left) and maximum vertical excursion (bottom left) vs. stellar mass for K2 stars with reliable ages. The distributions of $M_*$ (top left), $R_\mathrm{g}$ (top right) and $Z_\mathrm{max}$ (bottom right) are shown using kernel density estimates in bins of stellar age. The area under each distribution integrates to one.}
    \label{fig:Rg_Zmax_Mass}
\end{figure*}

It is also useful to investigate how these parameters change as a function of stellar age, as this can throw light on the dynamical evolution of the MW -- including radial migration and vertical heating. Figure \ref{fig:Rg_Zmax_Mass} shows $R_\mathrm{g}$ and $Z_\mathrm{max}$ as a function of stellar mass (which is more precisely constrained than age) and also the distributions of these parameters in six bins of age:
\begin{alignat*}{3}
                        \tau& < \qty[]{1}{Gyr},  &&      ~~\qty[]{1}{Gyr}  \leq \tau&& < \qty[]{2}{Gyr}, \\
    \qty[]{2}{Gyr} \leq \tau& < \qty[]{4}{Gyr},  &&      ~~\qty[]{4}{Gyr}  \leq \tau&& < \qty[]{6}{Gyr}, \\
    \qty[]{6}{Gyr} \leq \tau& < \qty[]{10}{Gyr}, &\quad&   \qty[]{10}{Gyr} \leq \tau&&.
\end{alignat*}
These age bins increase in width to accommodate the broader age posteriors of older stars. We see clearly that older stars (of lower mass) are distributed over a wider range of both $R_\mathrm{g}$ and $Z_\mathrm{max}$, and this spread is the signature of churning and vertical heating in our sample.

\begin{figure*}
    \centering
	\includegraphics[width=0.7\linewidth]{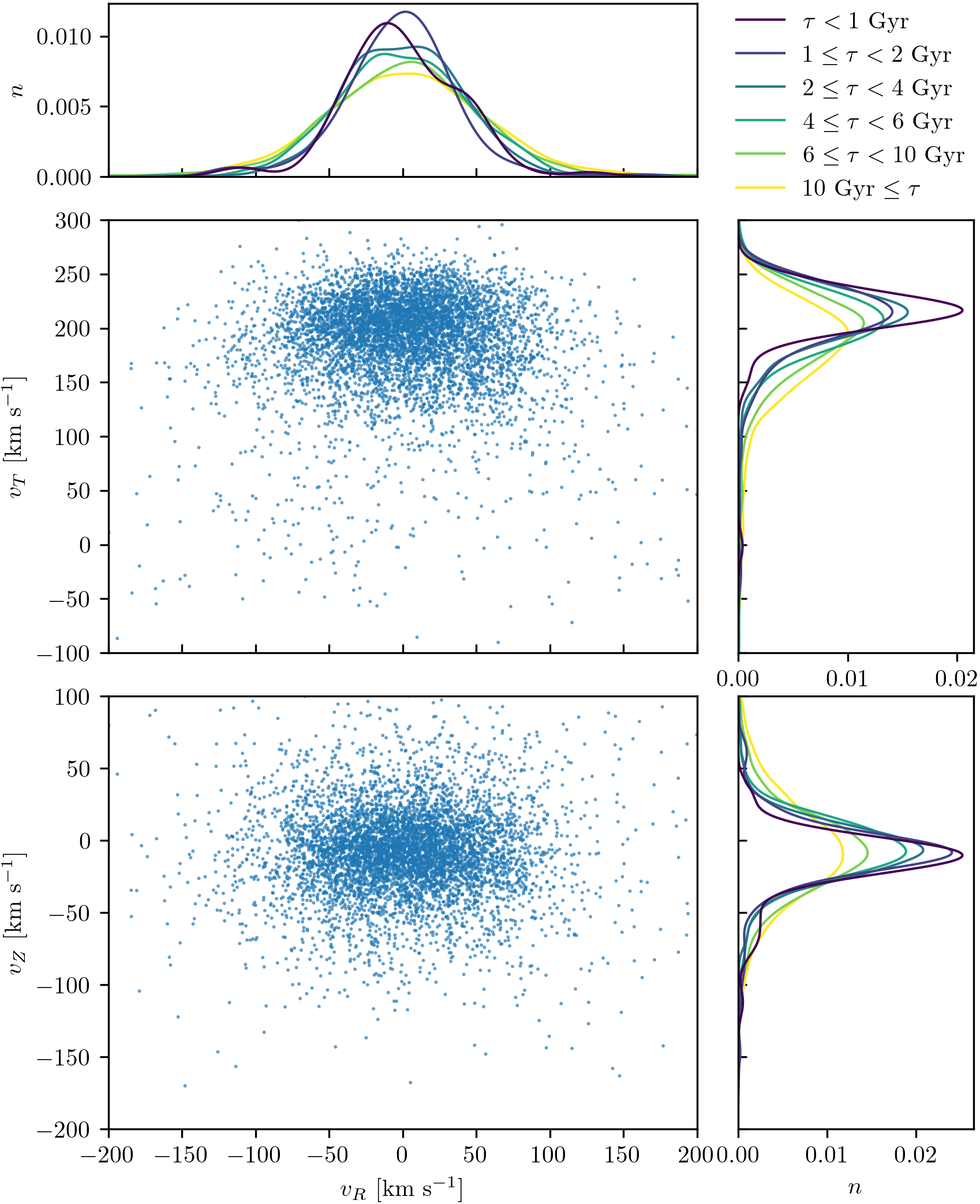}
    \caption{Tangential (middle left) and vertical (bottom left) velocity components vs. radial velocity component for K2 stars with reliable ages. The distributions of $v_\mathrm{R}$ (top left), $v_\mathrm{T}$ (top right) and $v_\mathrm{Z}$ (bottom right) are shown using kernel density estimates in bins of stellar age. The area under each distribution integrates to one.}
    \label{fig:vZ_vT_vR}
\end{figure*}

In Figure \ref{fig:vZ_vT_vR} we show a similar exercise, using the velocity components in Galactocentric cylindrical coordinates. We see similar trends with age to the previous plot, with older ages corresponding to a larger dispersion. This is a result both of the kinematic heating experienced by stars in the disc and the fact that the halo population is sampled only in the older bins. The vertical velocity dispersion $\sigma_{v_Z}$ has been studied as an indicator of the MW's last major merger, as it is expected that this event would have resulted in significant and rapid heating of the MW's disc at the time (e.g. \citealt{2018Natur.563...85H}; and see also \citealt{2020ARA&A..58..205H} and references therein). A simple visual inspection of the data in Figure \ref{fig:vZ_vT_vR} suggests that the largest change in $\sigma_{v_Z}$ between age bins is between the $4 - \qty[]{6}{\giga\year}{}$ and $6 - \qty[]{10}{\giga\year}{}$ bins (consider the width of the distributions in the bottom right panel), consistent with a merger \qty[]{10}{\giga\year}{} ago \citep[e.g.][]{2019NatAs...3..932G}. However, since the bin width also increases here from two to four Gyr, this indication should be treated with caution and further investigation is needed to confirm whether it can really be attributed to the MW's merger history.
 
\subsection{Chemical abundances and chemically selected populations}
\label{sec:app_chemistry}

\begin{figure*}
    \centering
	\includegraphics[width=0.9\linewidth]{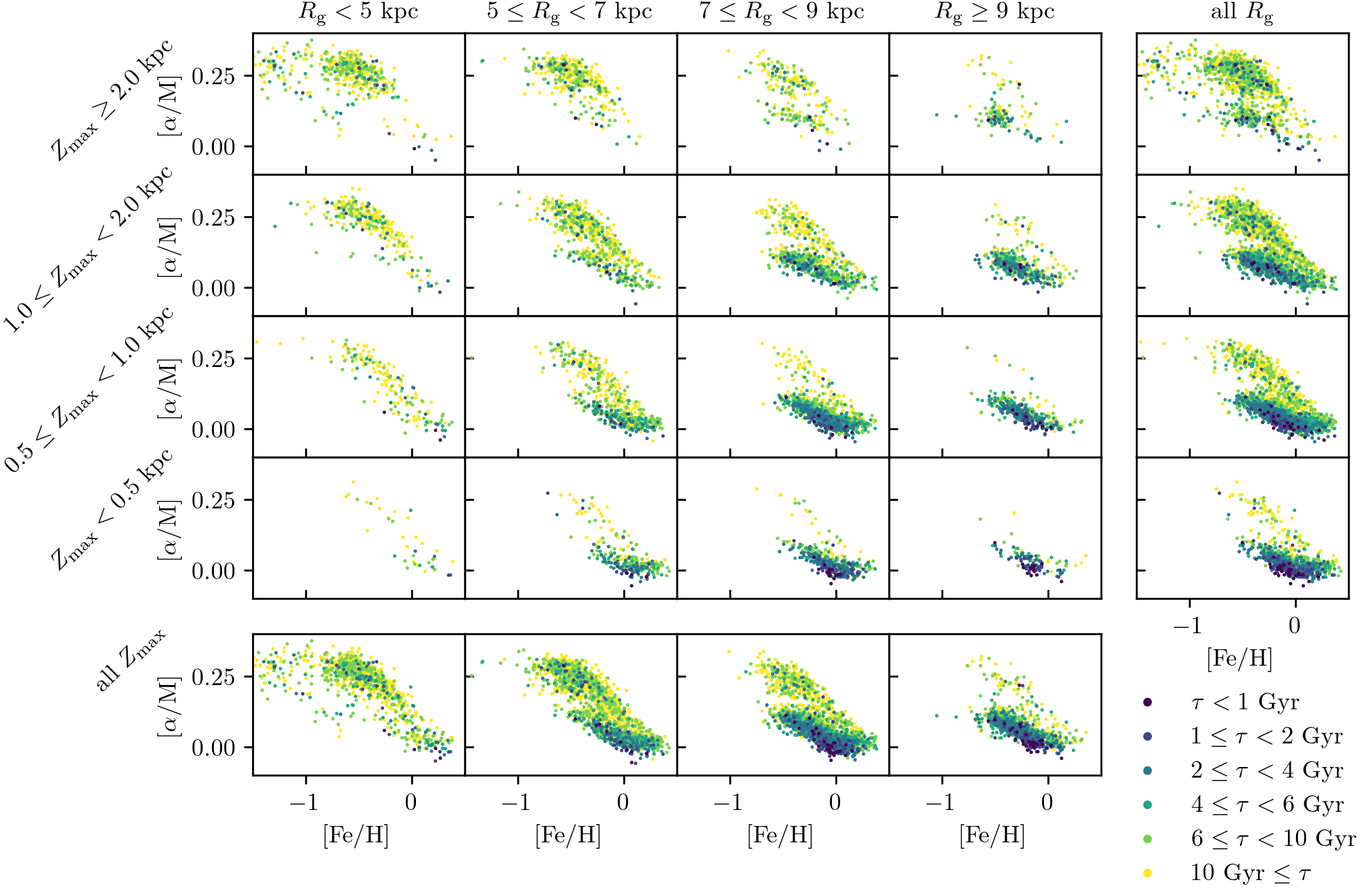}
    \caption{[$\alpha$/M] vs. [Fe/H] in bins of guiding radius and maximum vertical excursion (main panel; as shown in Figure \ref{fig:RgZmax}) for K2 stars with reliable ages. Also shown are the stars binned only by $R_\mathrm{g}$ (bottom panel) or $Z_\mathrm{max}$ (right panel). Stars are coloured according to age bin.}
    \label{fig:alpha_Fe_bins}
\end{figure*}

Moving beyond the kinematic properties, Figure \ref{fig:alpha_Fe_bins} shows the [$\alpha$/M] vs. [Fe/H] plane, in the bins of $R_\mathrm{g}$ and $Z_\mathrm{max}$ shown in Figure \ref{fig:RgZmax}. The stars are coloured according to their age bin, and the data are sorted to show the youngest stars on top. We see clearly that the low-$\alpha$ sequence is dominated by younger stars and the high-$\alpha$ sequence is formed primarily of stars older than \qty[]{6}{\giga\year}{}, but there are a few apparently young stars also present. These are interesting objects and are likely the result of binary evolution and mass accretion, rather than genuinely young \citep[e.g.][]{2015A&A...576L..12C, 2015MNRAS.451.2230M, 2018MNRAS.473.2984I, 2016A&A...595A..60J, 2023A&A...671A..21J}. \citet{2024A&A...683A.111G} studied these targets in more detail for this sample, while \citet{2021A&A...645A..85M} investigated those present in the \textit{Kepler} field, and both works support a scenario where these stars are products of mass transfer. Overall in Figure \ref{fig:alpha_Fe_bins}, we see a general trend of older, high-$\alpha$ stars in the inner disc, and younger stars concentrated further from the Galactic centre though, as also shown in Figure \ref{fig:Rg_Zmax_Mass}, old stars are found at all $R_\mathrm{g}$. We also see younger stars concentrated in the plane of the disc, while older stars tend to have orbits which extend to greater vertical distances \citep[though there are still old members of the thin disc visible in our data and e.g.][]{2024A&A...688A.167N}. This is consistent with the findings of other works in the literature (e.g. \citealt{2015ApJ...808..132H} Figure 4; \citealt{2023ApJ...954..124I} Figures 8 and 9).

\begin{figure}
    \centering
	\includegraphics[width=\linewidth]{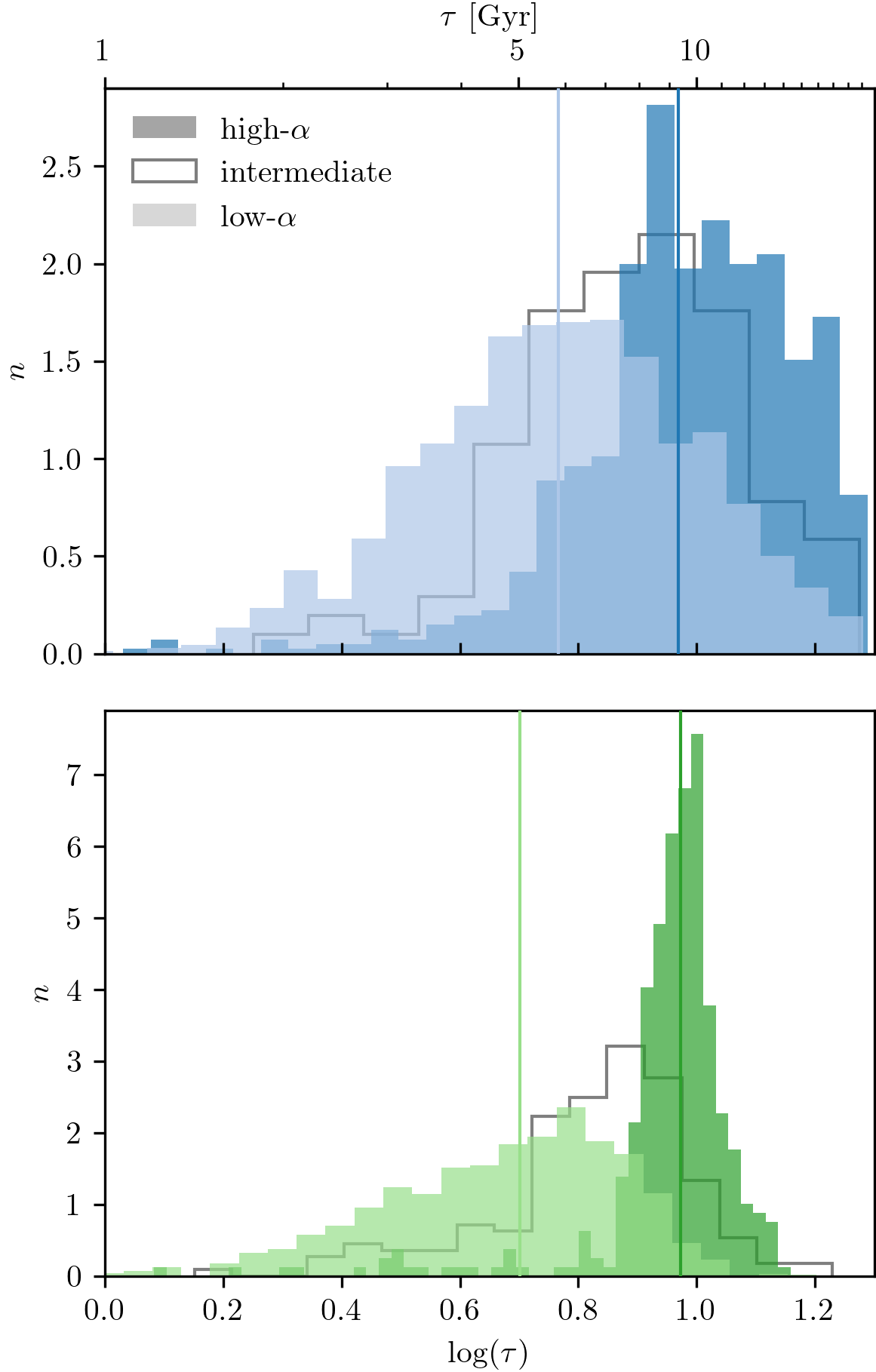}
    \caption{Distributions of logarithmic age for stars in the high- and low- and intermediate $\alpha$ regions of the [$\alpha$/M] vs. [Fe/H] plane (as shown in Figure \ref{fig:alpha_Fe}). The results are shown for likely-RGB stars from K2 (top) and \textit{Kepler} (bottom) with reliable ages (see Section \ref{sec:app_chemistry} for a description of the RGB selection). The area under each histogram integrates to one and the vertical lines show the median age of the low- and high-$\alpha$ populations.}
    \label{fig:alpha_agedists}
\end{figure}

In Figure \ref{fig:alpha_agedists} we show the age distributions of the high- and low-$\alpha$ sequences, as well as the `intermediate' stars from the higher metallicity region where the sequences appear to overlap (see Figure \ref{fig:alpha_Fe} for the separation of these sequences). We consider only RGB stars, as CHeB stars may have experienced significant mass-loss over their ascent of the RGB, resulting in additional systematic uncertainties in the age since the mass loss process is little understood. We select RGB stars using cuts in $L$ and $\log g$: 
\begin{equation*}
    L < \qty[]{35}{\solarlum}{}{}; ~\log g > 2.6
\end{equation*}
and CHeB stars:
\begin{equation*}
    L \geq \qty[]{35}{\solarlum}{}{}; ~\log g < 2.5,
\end{equation*}
where these selections are not continuous in $\log g$ to reduce cross-contamination between the populations. In both the \textit{Kepler} and K2 samples, the high-$\alpha$ population peaks at $\approx \qty[]{10}{\giga\year}{}$, the low-$\alpha$ at 6 - \qty[]{7}{\giga\year}{}, and the peak of the intermediate population is between the two. The median ages of these populations agree very well between \textit{Kepler} and K2, and the agreement is further improved when we remove the stars flagged as potentially affected too strongly by the age prior (as described in Section \ref{sssec:priors}). \citet{2021AJ....161..100W} compared ages from \textit{Kepler} and K2 asteroseismology and found that the high-$\alpha$ sequence in K2 was younger than in their \textit{Kepler} sample, but this is not the case in our data\footnote{ \citet{2021AJ....161..100W} separation of the high- and low-$\alpha$ sequences would have grouped our intermediate population with the low-$\alpha$ stars, so this cannot explain the difference in our results.} and is not present in the revised analysis of \citet{2024AJ....167..208W}.

\begin{figure}
    \centering
	\includegraphics[width=\linewidth]{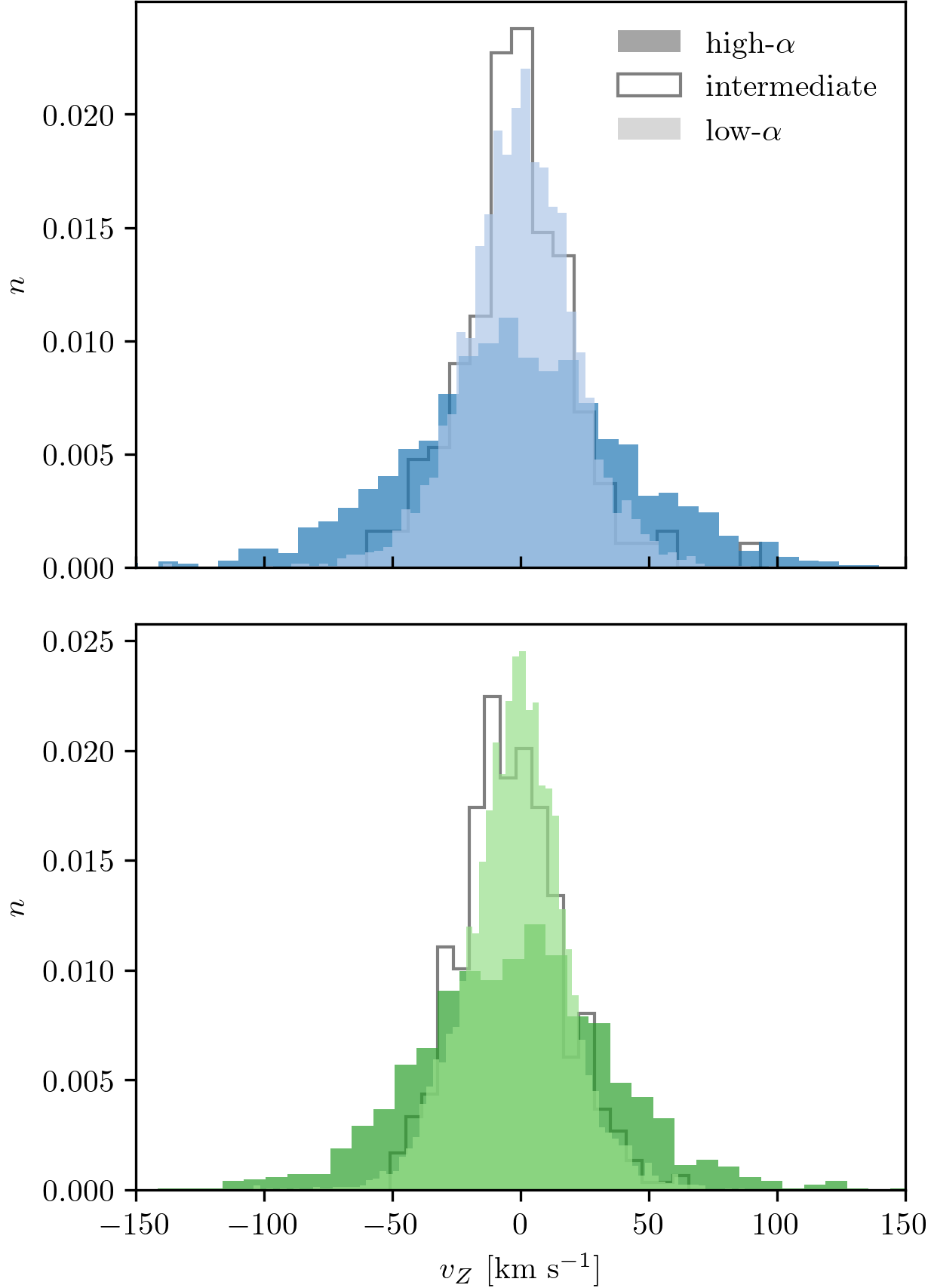}
    \caption{Distributions of vertical velocity for stars in the high- and low- and intermediate $\alpha$ regions of the [$\alpha$/M] vs. [Fe/H] plane (as shown in Figure \ref{fig:alpha_Fe}). The results are shown for likely-RGB stars from K2 (top) and \textit{Kepler} (bottom) with reliable ages (see Section \ref{sec:app_chemistry} for a description of the RGB selection). The area under each histogram integrates to one.}
    \label{fig:alpha_vzdists}
\end{figure}

Comparing the panels of Figure \ref{fig:alpha_agedists}, there appear to be many more high-$\alpha$ stars with ages greater than \qty[]{10}{\giga\year}{} present in the K2 sample than \textit{Kepler}. The qualitative differences between these two populations are in due to a combination of the larger uncertainties on the ages of the K2 targets (see Figure \ref{fig:age_err}) and their coverage of different Galactic components. Figure \ref{fig:alpha_vzdists} shows the $v_Z$ distributions of the same populations. In both \textit{Kepler} and K2, the low- and intermediate-$\alpha$ populations have very similar dispersions: $\sigma_{v_{Z}} \approx \qty[]{19}{\kilo\meter\per\second}{}$ in \textit{Kepler} and $\sigma_{v_{Z}} \approx \qty[]{22}{\kilo\meter\per\second}{}$ in K2. The high-$\alpha$ populations have a higher dispersion: $\sigma_{v_{Z}} \approx \qty[]{38}{\kilo\meter\per\second}{}$ in \textit{Kepler} and $\sigma_{v_{Z}} \approx \qty[]{44}{\kilo\meter\per\second}{}$ in K2. The larger velocity dispersions reflect the difference in the Galactic populations sampled by K2 (e.g. Figure \ref{fig:ELz} and \citealt{2018A&A...616A..11G} Figure 11) and supports the presence of at least some of these very old, kinematically hotter stars, but more precise age information is required to investigate this more robustly.

\begin{figure}
	\includegraphics[width=\linewidth]{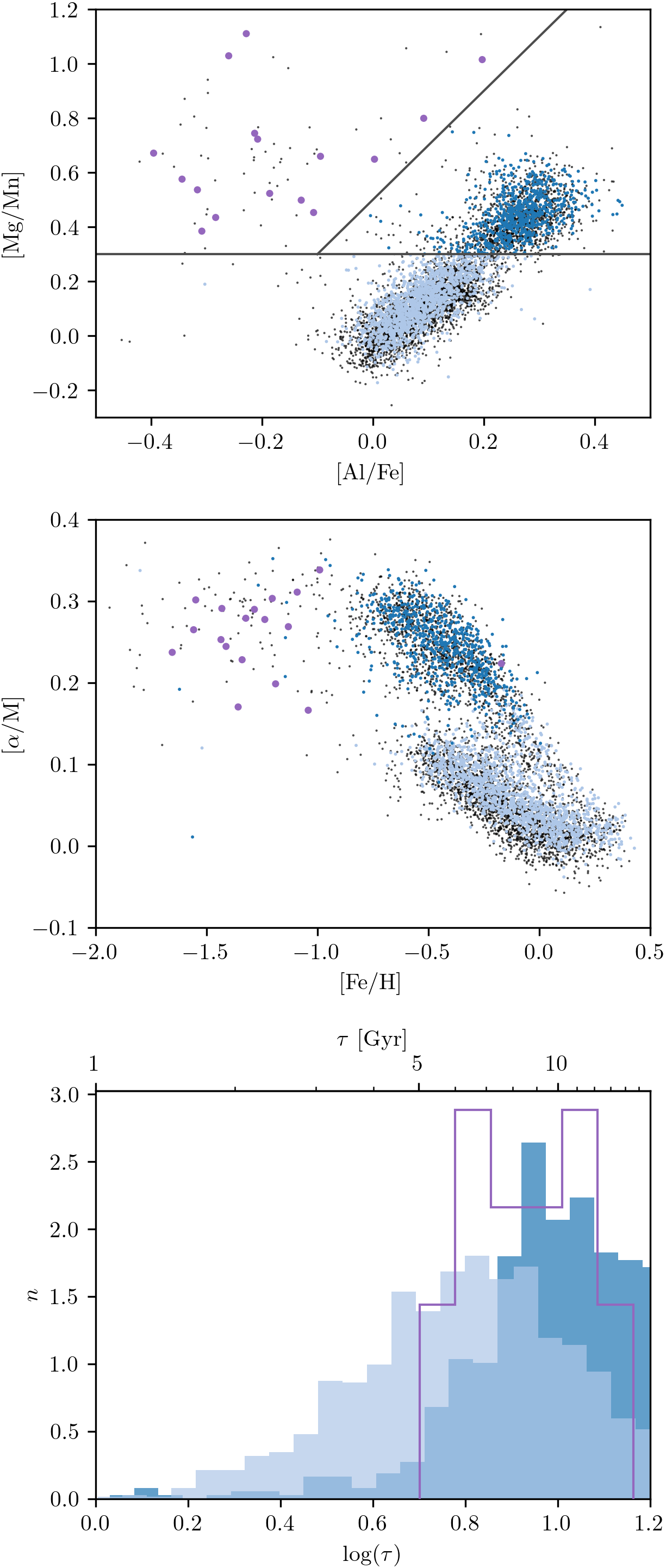}
    \caption{[Mg/Mn] vs. [Al/Fe] (top) and [$\alpha$/Fe] vs. [Fe/H] (middle) for K2 stars with reliable ages. Likely-RGB stars are shown in colour in the foreground (see Section \ref{sec:app_chemistry} for a description of the RGB selection) and the full sample is shown in the background (black points). Stars are divided, according to the grey lines in the top panel, into \textit{in situ} low-$\alpha$ (light blue), \textit{in situ} high-$\alpha$ (dark blue) and \textit{ex situ} (purple) populations. Also shown are the distributions of logarithmic age for these three populations (bottom).}
    \label{fig:exsitu}
\end{figure}

Figure \ref{fig:exsitu} shows a different separation of the K2 RGB sample and the corresponding age distributions. \citet{2015MNRAS.453..758H} and \citet{2020MNRAS.493.5195D} proposed the [Mg/Mn] vs. [Al/Fe] plane as a way of chemically identifying stars with a high probability of having been born outside the MW. These `\textit{ex situ}' stars have high [Mg/Mn] and low [Al/Fe] (top panel) and lie in the expected low-metallicity, high-$\alpha$ region of the [$\alpha$/M] vs. [Fe/H] plane (middle panel). We find that they are, on average, slightly younger than the `\textit{in situ}' high-$\alpha$ stars (bottom panel), with a median age of $\sim \qty[]{8.6}{\giga\year}{}$. This is consistent with literature expectations that the \textit{in situ} high-$\alpha$ population formed with higher star formation efficiency than the \textit{ex situ} \citep[e.g.][]{2014MNRAS.445.2575H, 2019NatAs...3..932G, 2019MNRAS.487L..47V, 2020MNRAS.493.5195D, 2021NatAs...5..640M}. To further refine the sample of \textit{ex situ} stars, we introduce kinematic information and consider stars with high eccentricity, $e$, as more likely to have been accreted \citep[e.g.][]{2019MNRAS.482.3426M}. From the chemically selected population of $18$ stars, $6$ have $e > 0.7$ with a median age of $\sim \qty[]{8.0}{\giga\year}{}$. The remaining $12$ stars (with $e \leq 0.7$) have a median age of $\sim \qty[]{8.8}{\giga\year}{}$. Repeating this analysis for the \textit{Kepler} sample, we find two stars with age $\sim \qty[]{8.8}{\giga\year}{}$.

\subsection{Signatures of internal mixing}
\label{sec:app_stellar}

\begin{figure*}
    \centering
	\includegraphics[width=0.9\linewidth]{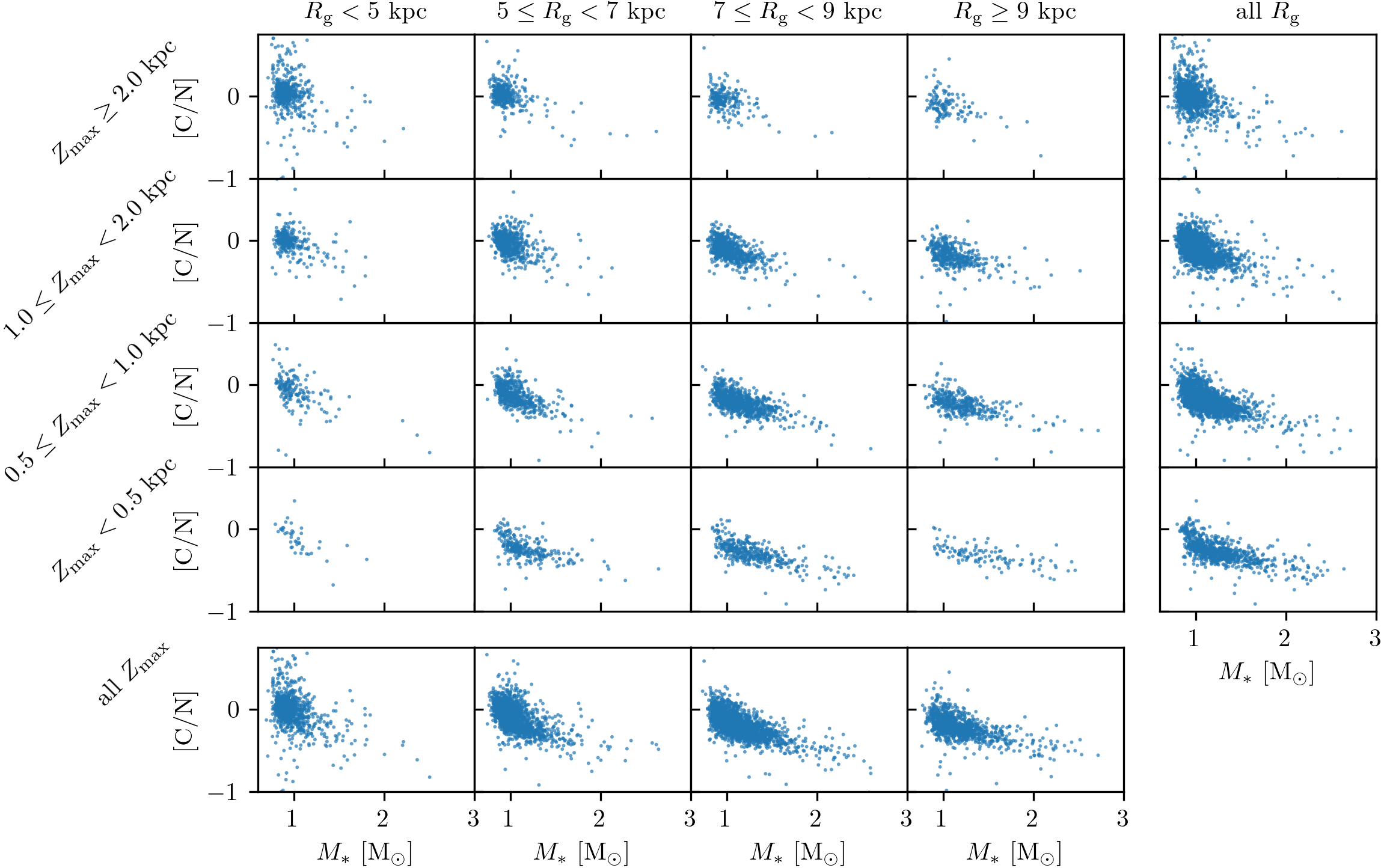}
    \caption{[C/N] vs. stellar mass in bins of guiding radius and maximum vertical excursion (main panel; as shown in Figure \ref{fig:RgZmax}) for K2 stars with reliable ages. Also shown are the stars binned only by $R_\mathrm{g}$ (bottom panel) or $Z_\mathrm{max}$ (right panel).}
    \label{fig:C_N_bins}
\end{figure*}

\begin{figure}
    \centering
	\includegraphics[width=\linewidth]{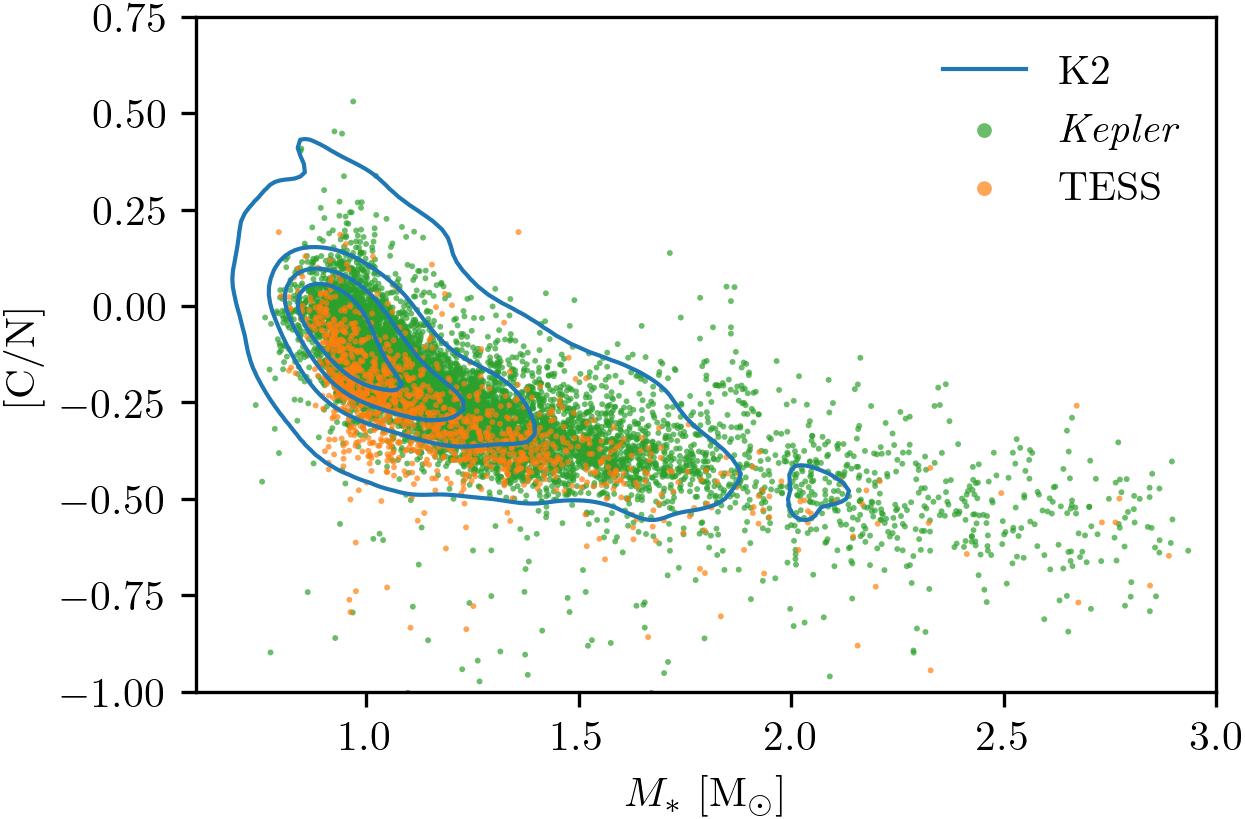}
    \caption{[C/N] vs. stellar mass for \textit{Kepler} (green points), TESS (orange points) and K2 (blue contours) stars with reliable ages.}
    \label{fig:C_N_sets}
\end{figure}

Results obtained with K2 are also consistent with \textit{Kepler} and TESS when we examine the [C/N] ratio as a function of stellar mass. The observed trend is a well-known consequence of the first dredge-up: by the end of the main sequence (MS) phase, the outer convection zone gradually extends deeper into the star, bringing to the surface material that has been partially processed by hydrogen burning during the MS phase. Moreover, for more massive stars, surface convection at the first dredge-up affects a larger portion of the star's total mass, leading to a decrease in the [C/N] ratio as the mass on the RGB increases \citep[e.g.][]{1964ApJ...140.1631I, 1967ApJ...147..624I, 2015MNRAS.453.1855M, 2016MNRAS.456.3655M}. Figure \ref{fig:C_N_bins} shows this relationship in the same bins of $R_\mathrm{g}$ and $Z_\mathrm{max}$ used previously, and we see changes across the Galaxy as a result of the different age demographics in each region. The increased scatter at low mass is likely due to variations in initial chemical composition for these old, metal-poor stars \citep[e.g.][]{1994PASP..106..553K, 2021arXiv210603912V, 2024MNRAS.530..149R}. The full K2 sample is shown by the contours in Figure \ref{fig:C_N_sets}, and maps well to the \textit{Kepler} and TESS samples.

\subsection{Chemical clocks}
\begin{figure*}
    \centering
	\includegraphics[width=0.9\linewidth]{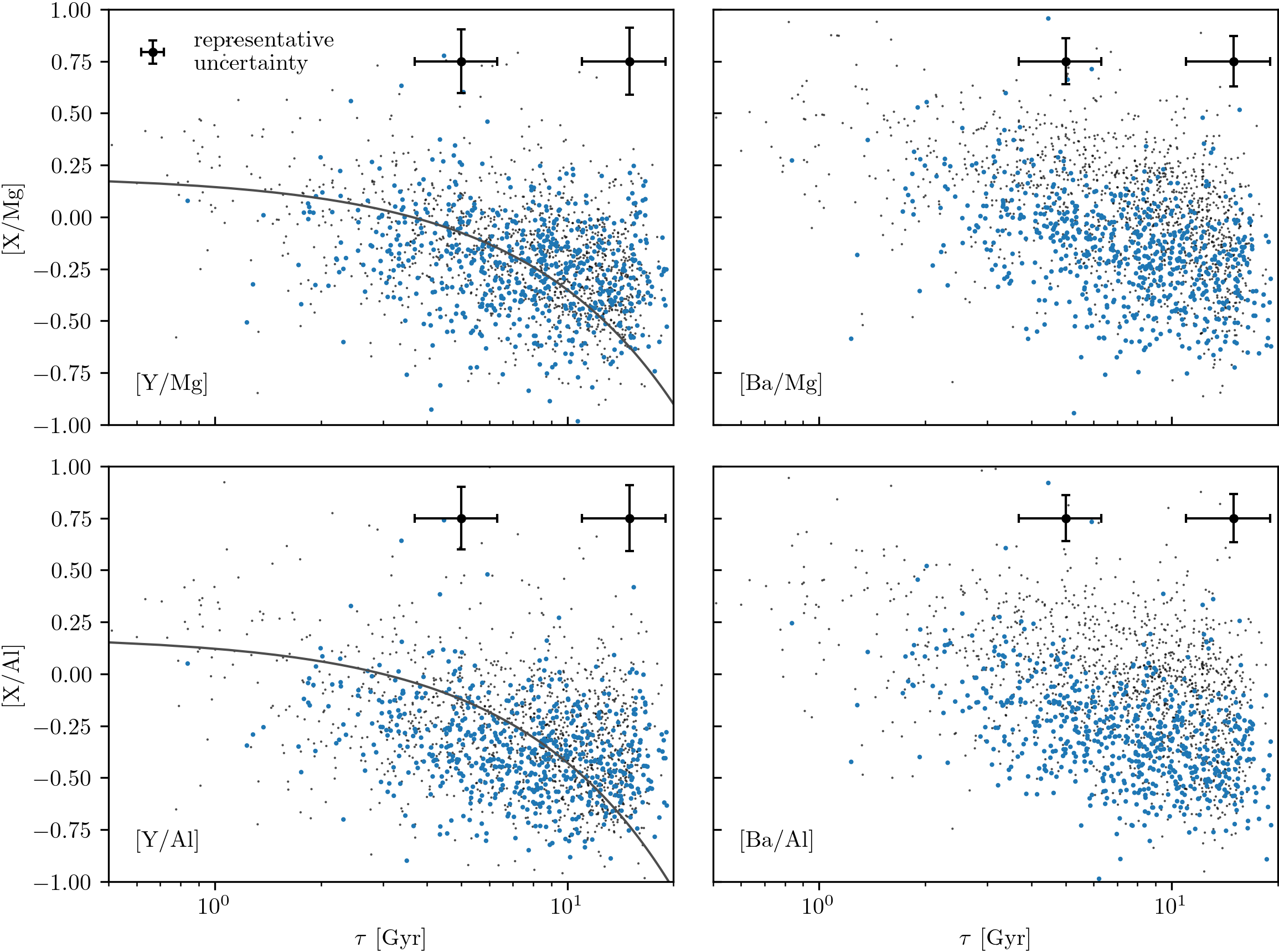}
    \caption{Abundance ratios vs. age for four elements commonly used in chemical clocks, from different nucleosynthetic channels: yttrium (left; light s-process), barium (right; heavy s-process), magnesium (top; $\alpha$ element) and aluminium (bottom; odd-Z element). The stars shown have reliable ages from the K2-GALAH cross-match and the likely-RGB sample is shown in blue points in the foreground. The black errorbars show median uncertainties on age an chemical abundance of stars younger (left point) or older (right point) than \qty[]{10}{\giga\year}{}. The grey line shows the relationship between the yttrium chemical clocks and age found by \citet{2021A&A...652A..25C}.}
    \label{fig:chem_clocks}
\end{figure*}

Finally, in Figure \ref{fig:chem_clocks}, we show four `chemical clocks' as a function of stellar age, from the K2-GALAH sample. Chemical clocks are formed by the ratio of two elements with different nucleosynthesis channels - ideally elements which individually have opposing trends with stellar age \citep[e.g., for RGs in Open Clusters see][]{2017A&A...604L...8S, 2020A&A...635A...8C, 2021A&A...652A..25C}. Identifying and testing consistent and sensitive chemical clocks is a valuable application of stellar ages, as they allow ages to be inferred for large populations of stars where asteroseismic or other age determinations are not available \citep{2022A&A...660A..15M}. \citet{2023A&A...677A..60C} completed a detailed study on chemical clocks using Ce and $\alpha$-elements in the K2-APOGEE sample, but introducing data from GALAH provides elements from neutron-capture nucleosynthesis channels. Here, we show chemical clocks formed by the s-process elements yttrium and barium (which are not available in APOGEE and have an increase in abundance relative to iron in older stars) compared to the $\alpha$-element magnesium and the odd-Z element aluminium (which have slightly decreasing trends relative to iron with age). Qualitatively our results agree with those reported in the literature \citep[e.g.][]{2021A&A...652A..25C}, and they demonstrate the utility of the GALAH sample, where elements from more nucleosynthesis channels can be investigated compared to APOGEE. However, though the uncertainties on age and chemical abundances are comparable, we see clearly that detailed analysis is limited by the precision on both axes.

\section{Summary and conclusions}
\label{sec:conclusion}

We have used asteroseismology based on data from \textit{Kepler}, K2 and TESS, combined with \textit{Gaia} DR3 astrometry and spectroscopic constraints from APOGEE DR17 and GALAH DR3 to infer stellar and orbital parameters for over $17,000$ RGs. We pay specific attention to the age estimates of our samples, identifying those which are or may be unreliable for different combinations of observational constraints. In particular, we highlight stars with unreliable $\Delta\nu$ determinations based on comparisons using \textit{Gaia} luminosities. These are particularly relevant in K2 data due to the short duration of the observations of each campaign, and therefore important to characterise for Galactic archaeology studies where the spatial range of K2 is a benefit.

We demonstrate the suitability of different age and mass estimations in different applications including showing trends with age and mass of the orbital parameters $R_\mathrm{g}$ and $Z_\mathrm{max}$ and velocity components, and investigating the [$\alpha$/M]-[Fe/H] plane in bins of $R_\mathrm{g}$, $Z_\mathrm{max}$ and age. We also compare the age distributions and vertical velocity dispersions of the low- and high-$\alpha$ populations, and the ages obtained from chemically selected \textit{ex situ} stars. 

We find that our K2 results agree with the known relationship between stellar mass and [C/N] ratio in \textit{Kepler} and TESS, but extend it to lower masses where the other surveys do not sample significantly. We observe greater scatter in this region, which we attribute to the greater spread of initial abundances amongst the oldest stars where the halo population is sampled. Finally, we use abundances from GALAH to show that our K2 ages agree qualitatively with results for chemical clocks based on open clusters.

These datasets, taken together, are a valuable tool for studies of Galactic archaeology and stellar evolution, and the K2 sample presented here has already been used to investigate open questions in the field of Galactic archaeology, such as the radial metallicity gradient \citep{2023MNRAS.526.2141W}, Galactic cerium enrichment \citep{2023A&A...677A..60C} and young $\alpha$-rich stars \citep{2024A&A...683A.111G}, while the \textit{Kepler} and K2 samples have been used to study mass loss during the red-giant-branch phase \citep{2024A&A...691A.288B}. 

The examples presented in these works represent a small fraction of the possible applications of these data which offer scope to investigate many open problems directly and to be used as training data which will extend the range of currently available machine learning-based ages.

Looking ahead, the availability of asteroseismology from the continuing TESS mission, as well as future missions including the Nancy Grace Roman Telescope \citep{2015arXiv150303757S} and ESA's PLATO mission \citep{2014ExA....38..249R}, will be complemented by future \textit{Gaia} data releases and upcoming spectroscopic surveys (e.g. 4MOST; \citealt{2012SPIE.8446E..0TD}, MOONS; \citealt{2014SPIE.9147E..0NC} and WEAVE; \citealt{2020SPIE11447E..14D}). PLATO, in particular, has the potential to provide age estimates with 10\% precision for large samples of RGs \citep{2017AN....338..644M}, which would represent a significant advancement over the present data.

\section*{Acknowledgements}
EW thanks the anonymous reviewer for their helpful comments, which improved the clarity of this paper.

EW, AM, KB, GC, VG, and AS acknowledge support from the ERC Consolidator Grant funding scheme (project ASTEROCHRONOMETRY, G.A. n. 772293 \url{http://www.asterochronometry.eu}).\\
SK is funded by the Swiss National Science Foundation through an Eccellenza Professorial Fellowship (award PCEFP2\_194638).\\
VG acknowledges financial support from INAF under the program “Giovani Astrofisiche ed Astrofisici di Eccellenza - IAF: Astrophysics Fellowships in Italy" (Project: GalacticA, "Galactic Archaeology: reconstructing the history of the Galaxy") and INAF Minigrant 2023.\\
AS has received funding from the European Research Council (ERC) under the European Union’s Horizon 2020 research and innovation programme (CartographY; grant agreement ID 804752).
DB acknowledges funding support by the Italian Ministerial Grant PRIN 2022, “Radiative opacities for astrophysical applications”, no. 2022NEXMP8, CUP C53D23001220006.

This paper includes data collected by the Kepler mission and obtained from the MAST data archive at the Space Telescope Science Institute (STScI). Funding for the Kepler mission is provided by the NASA Science Mission Directorate. STScI is operated by the Association of Universities for Research in Astronomy, Inc., under NASA contract NAS 5–26555.

This work has made use of data from the European Space Agency (ESA) mission {\it Gaia} (\url{https://www.cosmos.esa.int/gaia}), processed by the {\it Gaia} Data Processing and Analysis Consortium (DPAC, \url{https://www.cosmos.esa.int/web/gaia/dpac/consortium}). Funding for the DPAC has been provided by national institutions, in particular the institutions participating in the {\it Gaia} Multilateral Agreement.

Funding for the Sloan Digital Sky Survey IV has been provided by the Alfred P. Sloan Foundation, the U.S. Department of Energy Office of Science, and the Participating Institutions. 
SDSS-IV acknowledges support and resources from the Center for High Performance Computing  at the University of Utah. The SDSS website is \url{www.sdss4.org}.
SDSS-IV is managed by the Astrophysical Research Consortium for the Participating Institutions of the SDSS Collaboration including the Brazilian Participation Group, the Carnegie Institution for Science, Carnegie Mellon University, Center for Astrophysics | Harvard \& Smithsonian, the Chilean Participation Group, the French Participation Group, Instituto de Astrof\'isica de Canarias, The Johns Hopkins University, Kavli Institute for the Physics and Mathematics of the Universe (IPMU) / University of Tokyo, the Korean Participation Group, Lawrence Berkeley National Laboratory, Leibniz Institut f\"ur Astrophysik Potsdam (AIP),  Max-Planck-Institut f\"ur Astronomie (MPIA Heidelberg), Max-Planck-Institut f\"ur Astrophysik (MPA Garching), Max-Planck-Institut f\"ur Extraterrestrische Physik (MPE), National Astronomical Observatories of China, New Mexico State University, New York University, University of Notre Dame, Observat\'ario Nacional / MCTI, The Ohio State University, Pennsylvania State University, Shanghai Astronomical Observatory, United Kingdom Participation Group, Universidad Nacional Aut\'onoma de M\'exico, University of Arizona, University of Colorado Boulder, University of Oxford, University of Portsmouth, University of Utah, University of Virginia, University of Washington, University of 
Wisconsin, Vanderbilt University, and Yale University.

\textit{Software:} \texttt{astropy} \citep{2013A&A...558A..33A, 2018AJ....156..123A, 2022ApJ...935..167A}, \texttt{galpy} \citep{2015ApJS..216...29B}, \texttt{Matplotlib} \citep{2007CSE.....9...90H}, \texttt{NumPy} \citep{2020Natur.585..357H}, \texttt{pandas} \citep{reback2020pandas, mckinney-proc-scipy-2010}, \texttt{SciPy} \citep{2020NatMe..17..261V} and \texttt{TOPCAT} \citep{2005ASPC..347...29T}.




\section*{Data Availability}

The data underlying this article are available online at \url{https://doi.org/10.5281/zenodo.18417071}.



\bibliographystyle{mnras}
\bibliography{refs} 



\appendix

\section{Asteroseismic pipeline comparison for K2 targets}
\label{app:pipeline}

\begin{figure}
    \centering
	\includegraphics[width=\linewidth]{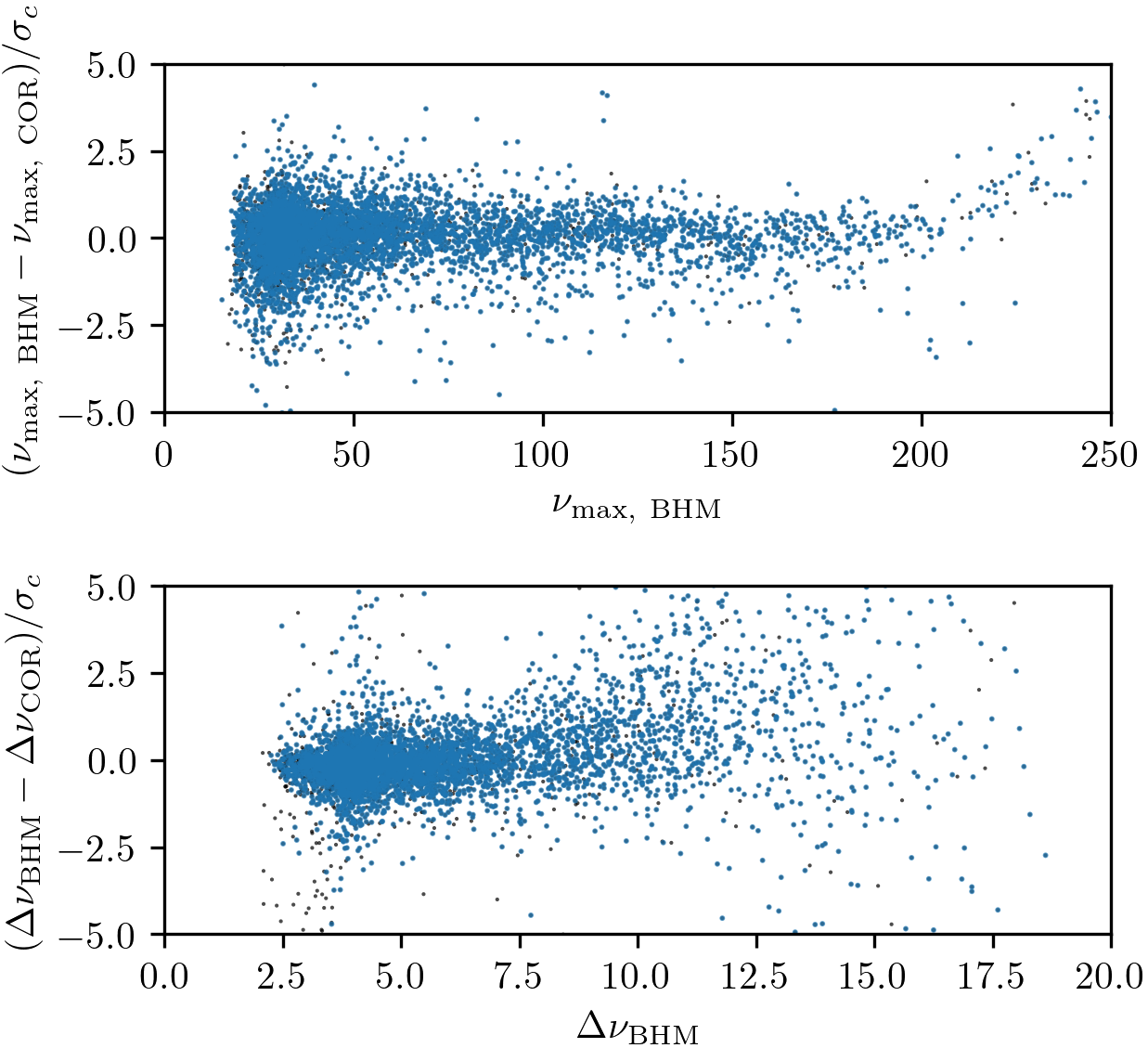}
    \caption{Difference between $\nu_\mathrm{max}$ (\textit{top}) and $\Delta \nu$ (\textit{bottom}) for K2 stars from BHM and COR pipelines, relative to combined uncertainty. Stars with reliable ages are shown in the foreground with blue points. Stars with reliable data but which do not pass our tests on age are shown as black points in the background. For clarity, we have not shown the few ($<1$ \% in $\nu_\mathrm{max}$ and $<5$ \% in $\Delta\nu$) stars where the disagreements exceed $5\sigma_c$.}
    \label{fig:numax_Dnu}
\end{figure}

Figure \ref{fig:numax_Dnu} shows a comparison of the asteroseismic parameters obtained for K2 stars from the independent BHM and COR pipelines, relative to their combined uncertainty. In general, the pipelines agree at the $2\sigma$ level.

\section{Identifying nyquist-affected $\nu_\mathrm{max}$}
\label{app:nyquist}

\begin{figure}
    \centering
	\includegraphics[width=\linewidth]{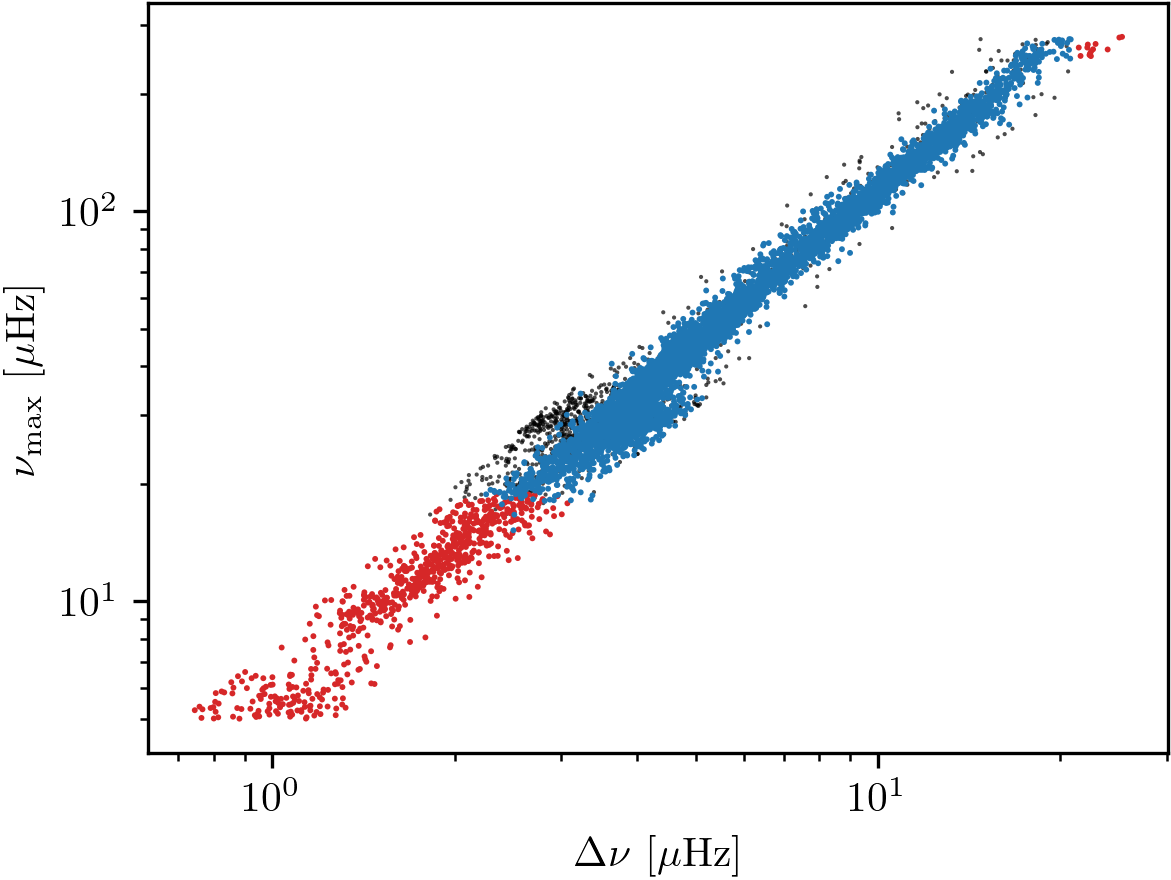}
    \caption{$\nu_\mathrm{max}$ vs. $\Delta\nu$ for the K2-BHM-APOGEE sample. Stars removed as a result of the cuts described in Section \ref{sec:seismo_cuts} are shown in red. Stars with reliable ages based on $\Delta\nu$ (see Section \ref{sssec:Dnu_L} for a description of the criteria) are shown in blue.}
    \label{fig:bad_numax}
\end{figure}

Figure \ref{fig:bad_numax} shows stars removed from the K2-BHM-APOGEE sample due to cuts on the asteroseismic parameters as red points. Targets with $\nu_\mathrm{max} + 3\sigma_{\nu_\mathrm{max}} < \qty[]{20}{\micro\hertz}{}{}$ and $\Delta\nu \geq \qty[]{21}{\micro\hertz}{}{}$ are rejected, as described in Section \ref{sec:seismo_cuts}.

\section{Catalogue Description}
The columns included in the catalogues released with this work are described in Table \ref{tab:cols}

\label{app:Cats}
\begin{table*}
\centering
\caption{Datamodel of the catalogues released with this work.}
\label{tab:cols}
\begin{tabular}{lllll}
Column Name & Suffixes & Details & Units &  \\ \cline{1-4}
\begin{tabular}[c]{@{}l@{}}\texttt{K2\_ID}/\\ \texttt{KIC\_ID}/\\ \texttt{TESS\_ID} \end{tabular} &  & K2/ \textit{Kepler}/ TESS identifier &  &  \\
\texttt{K2\_Campaign} &  & K2 campaign number (K2 Catalogues only) &  &  \\
\begin{tabular}[c]{@{}l@{}}\texttt{APOGEE\_ID}/\\ \texttt{GALAH\_ID}\end{tabular} &  & Spectroscopic survey identifier &  &  \\
\texttt{ASPCAP\_ID} &  & \begin{tabular}[c]{@{}l@{}}Identifies observation retained from APOGEE duplicates\\ (APOGEE catalogues only, see Section \ref{sssec:APG})\end{tabular} &  &  \\
\texttt{GaiaDR3\_ID} &  & \textit{Gaia} DR3 identifier &  &  \\ \cline{1-4}
\texttt{numax} & add \texttt{\_err} for uncertainty  & Frequency of maximum oscillation power & \qty[]{}{\micro\hertz}{} &  \\
\texttt{Dnu} & -`'- & Large frequency separation & \qty[]{}{\micro\hertz}{} &  \\
\texttt{EvoState} &  & \begin{tabular}[c]{@{}l@{}}From \citet{2018ApJS..236...42Y}. RGB = 1, HeB = 2, unclassified = 0 \\ ~~~~(\textit{Kepler} catalogue only) \end{tabular}&  &  \\ \cline{1-4}
\texttt{L} & add \texttt{\_err} for uncertainty & \begin{tabular}[c]{@{}l@{}}Luminosity based on \textit{Gaia} parallax \\ \qty[]{17}{\micro\hertz}{} zeropoint correction used in K2 catalogues and \\ Lindegren correction in \textit{Kepler} and TESS (see Section \ref{ssec:obs_gaia}) \end{tabular} & \qty[]{}{\solarlum}{} &  \\ \cline{1-4}
\texttt{Fe\_H} & add \texttt{\_err} for uncertainty & [Fe/H] from APOGEE or GALAH & dex &  \\
\begin{tabular}[c]{@{}l@{}}\texttt{alpha\_M}/\\ \texttt{alpha\_Fe}\end{tabular} & -`'- & [$\alpha$/M] from APOGEE or [$\alpha$/Fe] from GALAH & dex &  \\
\cline{1-4}
\texttt{Rg} & \begin{tabular}[c]{@{}l@{}}add \texttt{\_16} for 16th or \texttt{\_84} for 84th percentile then\\ add \texttt{\_BHM} or \texttt{\_Gaia} for source of distance \\ ~~~~(K2 BHM APOGEE catalogue only)\end{tabular} & Guiding radius in \texttt{MWPotential2014} & \qty[]{}{\kilo\parsec}{} &  \\
\texttt{Zmax} & -`'- & Maximum vertical excursion in \texttt{MWPotential2014} & \qty[]{}{\kilo\parsec}{} &  \\
\texttt{e} & -`'- & Orbital eccentricity in \texttt{MWPotential2014} &  &  \\
\texttt{Lz} & -`'- & Orbital angular momentum in \texttt{MWPotential2014} & \qty[]{}{\kilo\parsec\kilo\meter\per\second}{}&  \\
\texttt{E} & -`'- & Orbital energy in \texttt{MWPotential2014} & \qty[]{}{\kilo\meter\squared\per\second\squared}{} &  \\
\texttt{vR} & -`'- & Radial velocity component in \texttt{MWPotential2014} & \qty[]{}{\kilo\meter\per\second}{} &  \\
\texttt{vT} & -`'- & Tangential velocity component in \texttt{MWPotential2014} & \qty[]{}{\kilo\meter\per\second}{} &  \\
\texttt{vZ} & -`'- & Vertical velocity component in \texttt{MWPotential2014} & \qty[]{}{\kilo\meter\per\second}{} &  \\ \cline{1-4}
\texttt{age} & \begin{tabular}[c]{@{}l@{}}add \texttt{\_16} for 16th or \texttt{\_84} for 84th percentile then\\ add  \texttt{\_Dnu} or \texttt{\_L} for source of observational constraints\\ ~~~~(see Section \ref{sssec:Dnu_L})\end{tabular} & Age from PARAM & \qty[]{}{\giga\year}{} &  \\
\texttt{Mass} & -`'- & Mass from PARAM & \qty[]{}{\solarmass}{} &  \\
\texttt{Radius} & -`'- & Radius from PARAM & \qty[]{}{\solarradius}{} &  \\
\texttt{Dist} & -`'- & Distance from PARAM & \qty[]{}{\kilo\parsec}{} &  \\ \cline{1-4}
\texttt{warn\_low\_numax} & & ${\nu_\mathrm{max}}$ more than three standard deviations below \qty[]{20}{\micro\hertz}{} &  &  \\
\texttt{warn\_high\_Dnu} & & $\Delta\nu \geq \qty[]{21}{\micro\hertz}{}$ &  &  \\
\texttt{warn\_high\_RUWE} &  & from \textit{Gaia} DR3 &  &  \\
\texttt{warn\_NSS} &  & -`'- &  &  \\
\begin{tabular}[c]{@{}l@{}}\texttt{warn\_APOGEE}/\\ \texttt{warn\_GALAH}\end{tabular} &  & Combination of flags from spectroscopy (see Section \ref{ssec:obs_spectro}) &  &  \\
\texttt{drop\_K2\_dup} & & Duplicate target removed (K2 catalogues only) &  & \\
\texttt{warn\_plax\_err} &  & $\varpi / \sigma_\varpi < 5$ &  &  \\
\texttt{warn\_dustmaps} &  & Validity of dustmap outputs (\textit{Kepler} and K2 catalogues only) &  &  \\
\texttt{warn\_nveAv} & add \texttt{\_Dnu} or \texttt{\_L} for source of observational constraints & PARAM returns negative extinction &  &  \\
\texttt{warn\_massDiff} &  & \begin{tabular}[c]{@{}l@{}} Difference in mass from scaling relations based on $L$ or $\Delta\nu$ \\ exceeds 50\% (See Section \ref{sssec:Dnu_L})\end{tabular} &  &  \\
\texttt{Rel\_data} & add \texttt{\_Dnu} or \texttt{\_L} for source of observational constraints & \begin{tabular}[c]{@{}l@{}} combines flags from \texttt{warn\_low\_numax} to \texttt{drop\_K2\_dup} \\ for $\Delta\nu$ datasets \\ or from \texttt{warn\_low\_numax} to  \texttt{warn\_dustmaps} for $L$ datasets \\ to return stars with reliable data \end{tabular}&  &  \\
\texttt{Rel\_age} & -`'- & \begin{tabular}[c]{@{}l@{}} combines \texttt{Rel\_data} with \texttt{warn\_nveAv} and \texttt{warn\_massDiff} \\ to return stars with reliable ages \end{tabular} &  &  \\

\end{tabular}
\end{table*}

\section{Comparing $\Delta\nu$ and $L$ results in \textit{Kepler} and TESS}
\label{app:M_R_COR}

\begin{figure}
    \centering
	\includegraphics[width=\linewidth]{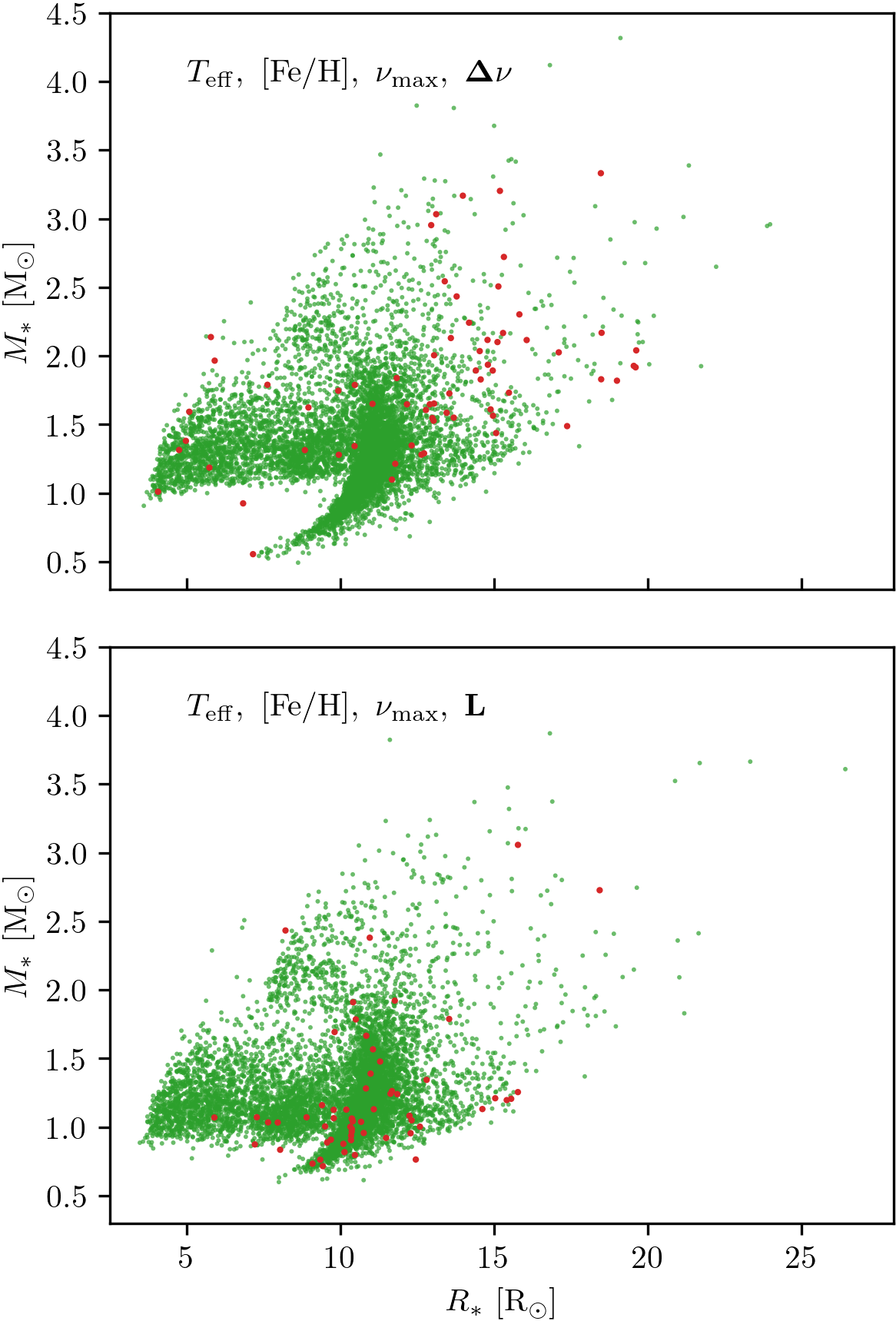}
    \caption{Same as Figure \ref{fig:M_R}, but for \textit{Kepler} targets.}
    \label{fig:M_R_Kepler}
\end{figure}

\begin{figure}
    \centering
	\includegraphics[width=\linewidth]{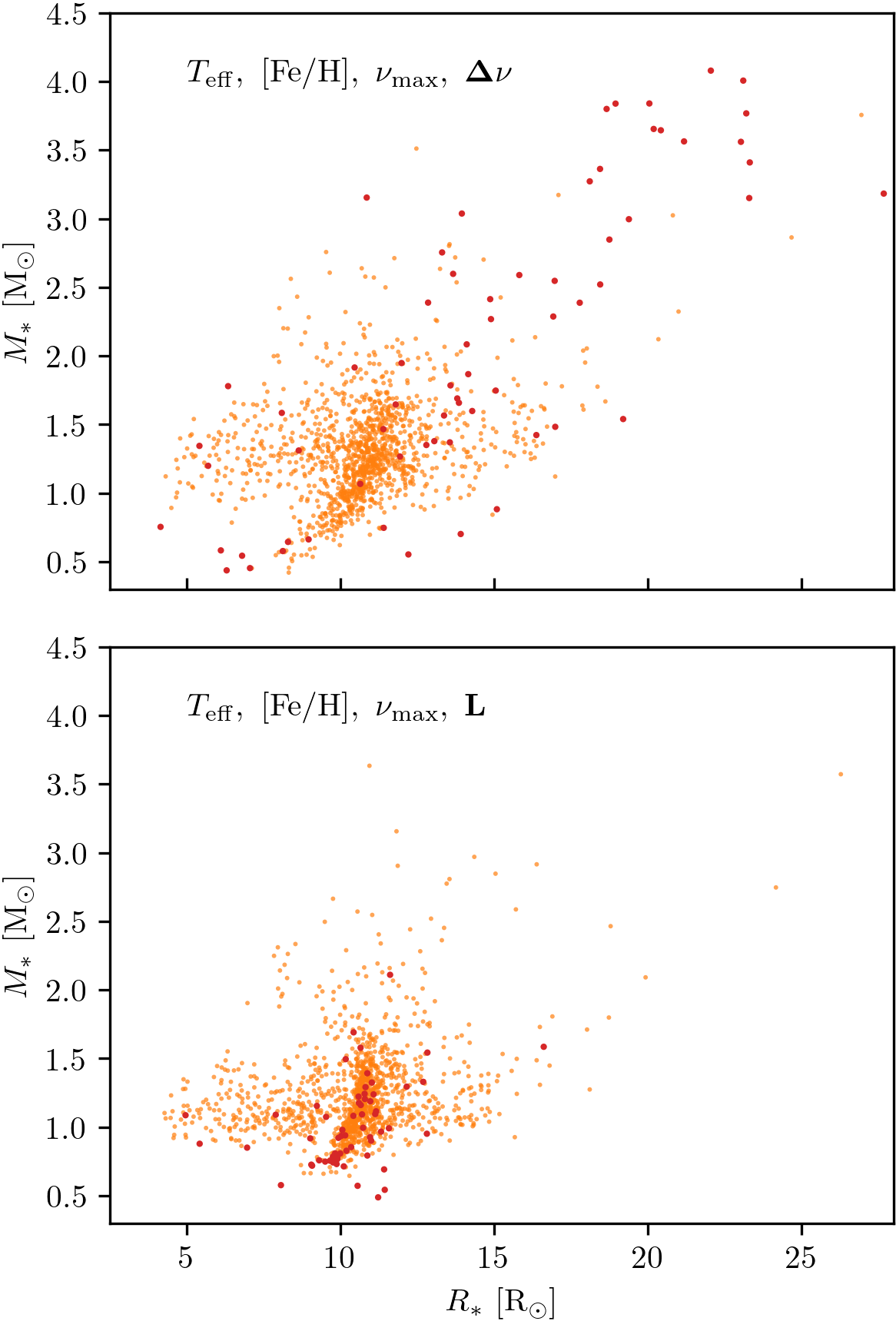}
    \caption{Same as Figure \ref{fig:M_R}, but for TESS targets.}
    \label{fig:M_R_TESS}
\end{figure}

Figures \ref{fig:M_R_Kepler} and \ref{fig:M_R_TESS} show the stellar mass and radius obtained from scaling relations for \textit{Kepler} and TESS stars, with stars flagged as having unreliable $\Delta\nu$ according to the definition in Section \ref{sssec:Dnu_L} highlighted (red points). In comparison to Figure \ref{fig:M_R}, we see a greatly reduced number of stars highlighted by this flag, as a result of the increased duration of the observations. In K2, 782 stars with otherwise reliable data are removed on the basis of this cut (11\%), in \textit{Kepler} and TESS 66 (0.8\%) and 72 (5\%) stars are removed, respectively.

\section{Comparison of ages to the APO-K2 catalogue}
\label{app:APO-K2_comp}

\begin{figure}
    \centering
	\includegraphics[width=\linewidth]{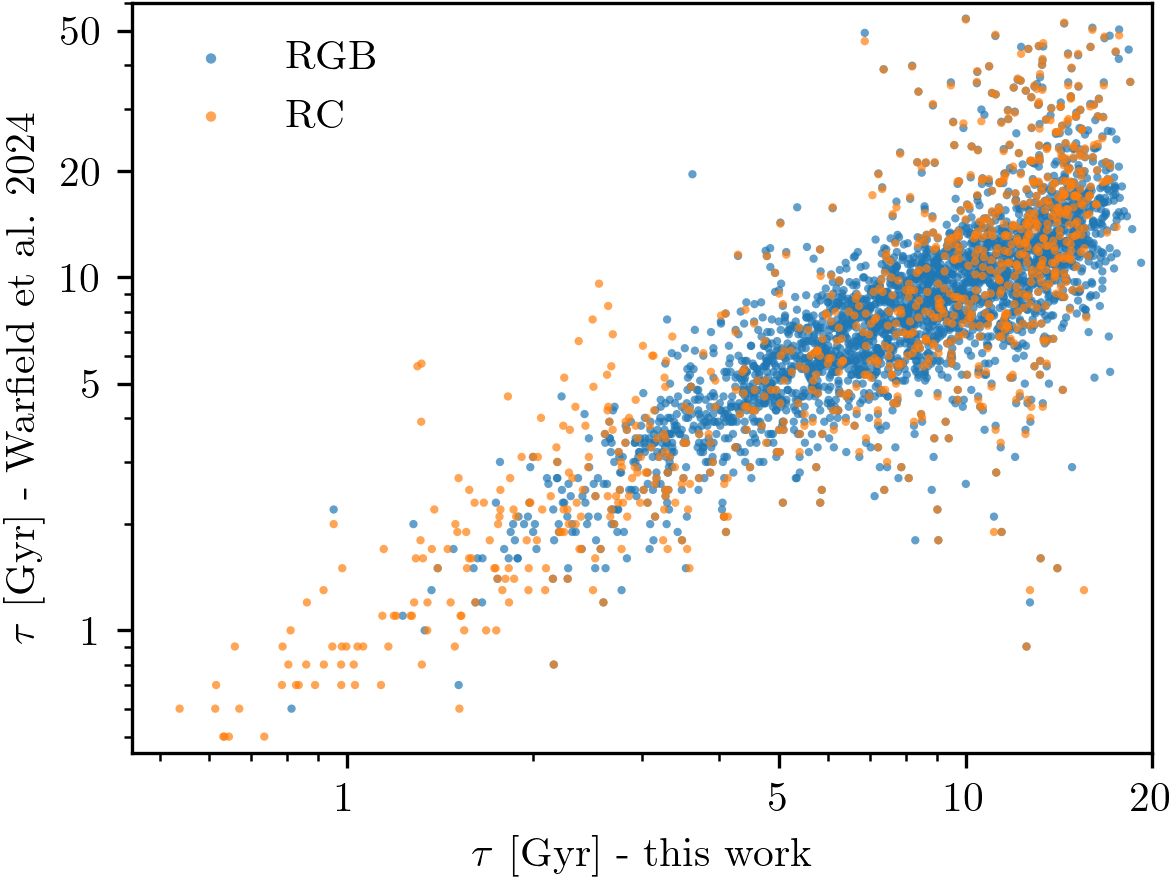}
    \caption{Comparison of ages for stars in common between this work and the APO-K2 catalogue. Evolutionary states are taken from \citet{2024AJ....167..208W}.}
    \label{fig:Warf1}
\end{figure}

\begin{figure}
    \centering
	\includegraphics[width=\linewidth]{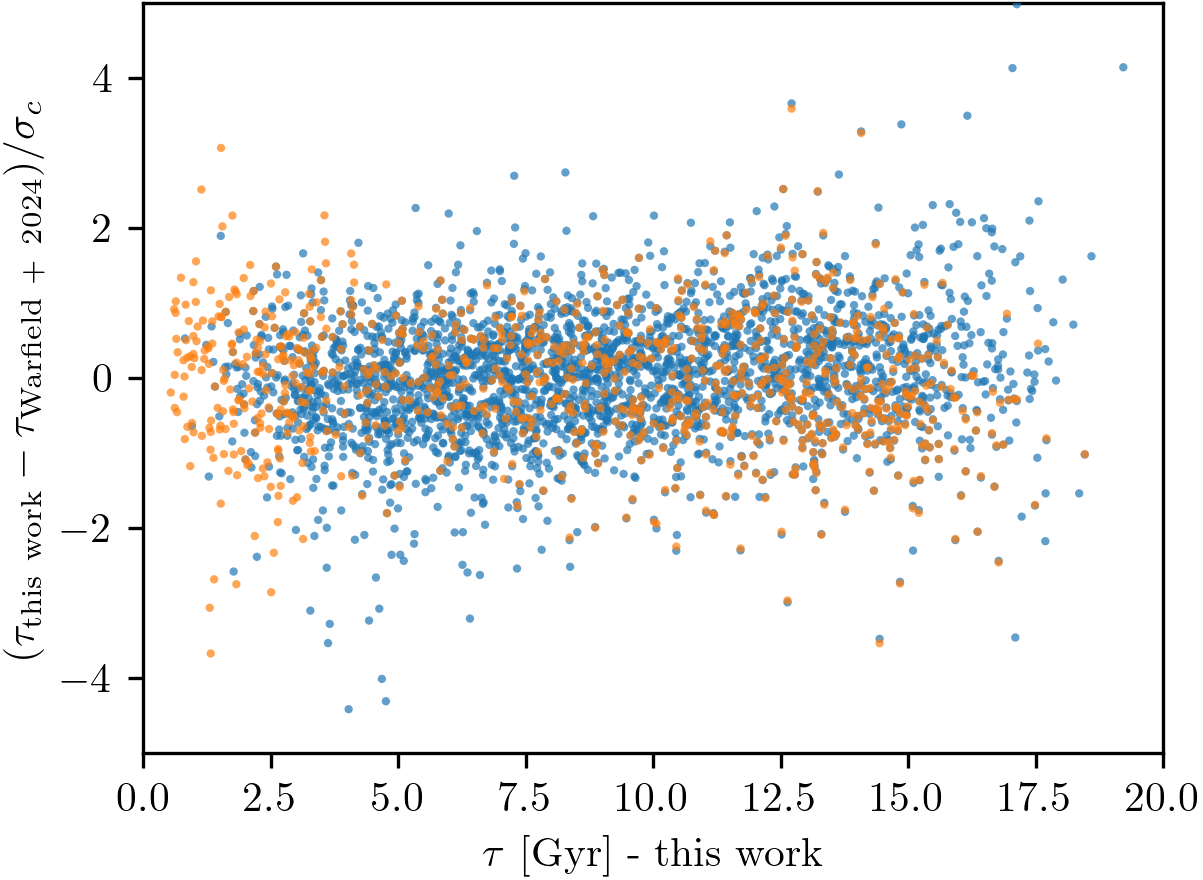}
    \caption{Difference between the ages presented in this work and those from APO-K2, relative to combined uncertainty. The colours are the same as Figure \ref{fig:Warf1}. For clarity, three stars which agree at the $5~-~15\sigma$ level are not shown.}
    \label{fig:Warf2}
\end{figure}

\citet{2024AJ....167..208W} presented ages for 4661 RGB and a further 1912 RC stars, with evolutionary states based on spectroscopy, described in \citet{2021AJ....161..100W}. In Figures \ref{fig:Warf1} and \ref{fig:Warf2}, we show a comparison of the reliable ages constrained by $[T_\mathrm{eff}, \mathrm{[Fe/H]}, \nu_\mathrm{max}, \Delta\nu]$ from the BHM-APOGEE sample and those from APO-K2. There are 2989 stars in common with the APO-K2 RGB sample, and 939 in the RC.

\bsp	
\label{lastpage}
\end{document}